\documentclass[onecolumn]{aastex}
\usepackage{emulateapj5}
\usepackage{apjfonts}
\usepackage{epsf}
\slugcomment{ApJ, 599, 7-23 (2003)}
\shorttitle{ARC STATISTICS IN TRIAXIAL DARK MATTER HALOS}
\shortauthors{OGURI, LEE, \& SUTO}

\begin{document}
\title{Arc Statistics in Triaxial Dark Matter Halos: \\
Testing the Collisionless Cold Dark Matter Paradigm}
%
\author{Masamune Oguri, Jounghun Lee,
and Yasushi Suto\altaffilmark{1}}
\affil{Department of Physics, School of Science, University of Tokyo, 
 Tokyo 113-0033, Japan}
\email{oguri@utap.phys.s.u-tokyo.ac.jp, lee@utap.phys.s.u-tokyo.ac.jp,
suto@phys.s.u-tokyo.ac.jp} 
\altaffiltext{1}{Also at 
Research Center for the Early Universe (RESCEU), School
of Science, University of Tokyo, Tokyo 113-0033, Japan}
%
%
\begin{abstract}
Statistics of lensed arcs in clusters of galaxies serve as a powerful
probe of both the non-sphericity and the inner slope of dark matter
halos.  We develop a semi-analytic method to compute the number of arcs
in triaxial dark matter halos. This combines the lensing cross section
from the Monte Carlo ray-tracing simulations, and the probability
distribution function (PDF) of the axis ratios evaluated from
cosmological $N$-body simulations.  This approach enables one to
incorporate both asymmetries in the projected mass density and
elongations along the line-of-sight analytically, for the first time in
cosmological lensed arc statistics.  As expected, triaxial dark matter
halos significantly increase the number of arcs relative to spherical
models; the difference amounts to more than one order of magnitude while
the value of enhancement depends on the specific properties of density
profiles. Then we compare our theoretical predictions with the observed
number of arcs from $38$ X-ray selected clusters. In contrast to the
previous claims, our triaxial dark matter halos with inner density
profile $\rho \propto r^{-1.5}$ in a Lambda-dominated cold dark matter
(CDM) universe reproduces well the observation. Since both the central
mass concentration and triaxial axis ratios (minor to major axis ratio
$\sim 0.5$) required to account for the observed data are consistent
with cosmological N-body simulations, our result may be interpreted to
lend strong support for the collisionless CDM paradigm at the mass scale
of clusters.
\end{abstract} 

%
%
\section{Introduction}

The discovery of a lensed arc in a rich cluster A370
\citep{lynds86,soucail87} opened a direct window to probe the dark mass
distribution in clusters of galaxies.  Since gravitational lensing
phenomena are solely dictated by intervening mass distributions, they
are not biased by the luminous objects unlike other conventional
observations.  Indeed, previous work \citep*{wu93,miralda93a,miralda95,
miralda02,bartelmann96,hattori97b,molikawa99,williams99,meneghetti01,
molikawa01,oguri01,sand02,gavazzi03} showed that the number, shapes, 
and positions of lensed arcs are sensitive to the mass distribution of
clusters.  For instance, \citet{oguri01} calculated the number of arcs
using the generalization of the universal density profile proposed by
\citet*{navarro96,navarro97} and pointed out that it is extremely sensitive
to the inner slope and the concentration parameter of the density
profile; the number of arcs changes by more than an order of magnitude
among different models that are of cosmological interest. Therefore 
lensing arc surveys provide an important probe of density profiles of
clusters in a complementary manner to the statistics of wide-separation 
lensed quasars \citep{maoz97,keeton01c,oguri02b,oguri03,li03}.

While most previous studies of lensed arcs have aimed at constraining
the cosmological parameters
\citep*{wu96,bartelmann98,cooray99,sereno02,golse02,bartelmann03}, we
rather focus on extracting information of the density profiles of dark
matter halos. Thus we assume a Lambda-dominated cold dark matter (CDM) model
that is consistent with the recent {\it Wilkinson Microwave Anisotropy Probe}
({\it WMAP}) result \citep{spergel03}; the matter
density parameter $\Omega_0=0.3$, the dimensionless cosmological
constant $\lambda_0=0.7$, the mass fluctuation amplitude $\sigma_8=0.9$,
the Hubble constant in units of $100{\rm km\,s^{-1}Mpc^{-1}}$,
$h=0.7$. In fact, arc statistics depend on the assumed set of
cosmological parameters in two ways; directly through the geometry of
the universe and somewhat indirectly through properties of density
profiles which also depend on the cosmology. For instance,
\citet{bartelmann98} found that the numbers of arcs significantly change
among different cosmological models, and concluded that only open CDM
models can reproduce the high frequency of observed
arcs. \citet{oguri01} showed, however, that the result largely comes
from the larger concentration parameter of halo profiles in the open CDM
model than in the Lambda-dominated CDM model. Thus this may be more
related to the small-scale behavior of the CDM model than the ``global''
effect of the cosmological constant.

The values of cosmological parameters are determined fairly
accurately now, thus our primary interest here is to confront the density
profiles of dark matter halos with the arc statistics, and thereby we
would like even to test the collisionless CDM paradigm.  For this
purpose, a non-spherical description for the lensing halos is the most
essential since cross sections for arcs are quite sensitive to the
non-sphericity of mass distribution
\citep*[e.g.,][]{bartelmann95a,bartelmann95b,meneghetti01,oguri02a}. 
Indeed, previous analytic models
adopting spherical lens models failed to reproduce the observed high
frequency of arcs \citep{hattori97b,molikawa99}. Because of the
lack of a realistic analytical model for non-spherical lens, however,
one had to resort to N-body simulations to take account of non-spherical
effects on the arcs statistics \citep[e.g.,][]{bartelmann94}.
Nevertheless it is quite demanding for those numerical simulations to
resolve the central part of the gravitational potential of lensing
halos while keeping the reasonable number of those objects sufficient
for statistical discussion. This is why a complementary (semi-)
analytical approach to the arc statistics is highly desired.

Recently, new methods to constrain the mass profile of individual
clusters also have been developed
\citep{smith01,sand02,clowe01,clowe02,gavazzi03}. Although such methods
can measure mass distributions of individual clusters precisely, it
may suffer from the special selection function and the scatter around
the mean mass distribution. For instances, analysis of clusters only
with giant arcs  may result in more elongated clusters than average
because JS02 showed that triaxial axis ratios have fairly broad
distributions. Therefore it is of great importance to study statistics
of lensed arcs which allow us to obtain information on the mean profile. 

In this paper, we develop and study in detail, for the first time, such
an analytical model of the non-spherical lensing objects for the arc
statistics. Specifically we adopt the triaxial description of dark
matter halos proposed by \citet[hereafter JS02]{jing02}. They have
presented detailed triaxial modeling of halo density profiles, which
enables us to incorporate the asymmetry of dark matter halos
statistically and systematically.  We first compute the lensing cross
sections for arcs on the basis of the Monte Carlo simulations following
\citet{oguri02a}. Then we make systematic predictions of the number of
arcs by averaging the cross sections over the probability distribution
functions (PDFs) of the axis ratios and the concentration parameters
and assuming the random orientation of the dark halos along the
line-of-sight of the observer. Those theoretical predictions are
compared with the number of observed arcs in a sample of $38$ X-ray
selected clusters compiled by \citet{luppino99}.  We pay particular
attention to several selection functions of clusters and arcs which may
systematically affect our results \citep*[e.g.,][]{wambsganss03}.

The plan of this paper is as follows. In \S \ref{sec:halo}, we briefly
summarize the triaxial modeling proposed by JS02. We present several key
results of gravitational lensing in triaxial dark matter halos in \S
\ref{sec:lens}. The method to predict the number of arcs is described in
detail in \S \ref{sec:arcstat}, and comparison with observations is
discussed in \S \ref{sec:obs}.  Finally, we discuss several implications
of our results in \S \ref{sec:discussion}, and summarize the conclusion
in \S \ref{sec:conclusion}.

\section{Description of Triaxial Dark Matter Halos\label{sec:halo}}

In this section, we briefly summarize the triaxial model of dark matter
halos proposed by JS02.  They obtained the detailed triaxial modeling on
the basis of their high-resolution individual halo simulations as well
as large-scale cosmological simulations. Most importantly, they provided
a series of useful fitting formulae for mass- and redshift-dependence
and the PDFs of the axis ratio and the concentration parameter.  Such
detailed and quantitative modeling enables us to incorporate the
non-sphericity of dark matter halos in a reliable manner.

\subsection{Coordinate Systems}

We introduce two Cartesian coordinate systems, $\vec{x} = (x,y,z)$ and 
$\vec{x'} = (x',y',z')$, which represent respectively the principal
coordinate system of the triaxial dark halo and the observer's 
coordinate system. The origins of both coordinate systems are
set at the center of the halo.  It is assumed that the $z'$-axis 
runs along the line-of-sight direction of the observer, and that the
$z$-axis lies along the major principal axis.  In general, the
relative orientation between the two coordinate systems can be specified
by the three Euler angles. However, in our case, it is only the
line-of-sight direction that is fixed while the rotation angle of the
$x'$-$y'$ plane relative to $x$-$y$ plane is arbitrary, and thus we may need
only two angles to specify the relative orientation of the two
coordinate systems. Here we make a choice of $x'$-axis lying in the
$x$-$y$ plane. Then the relative orientation of the two coordinate systems
can be expressed in terms of the line-of-sight direction in the halo 
principal coordinate system. 

\begin{figure*}[t]
\epsscale{0.5}
\plotone{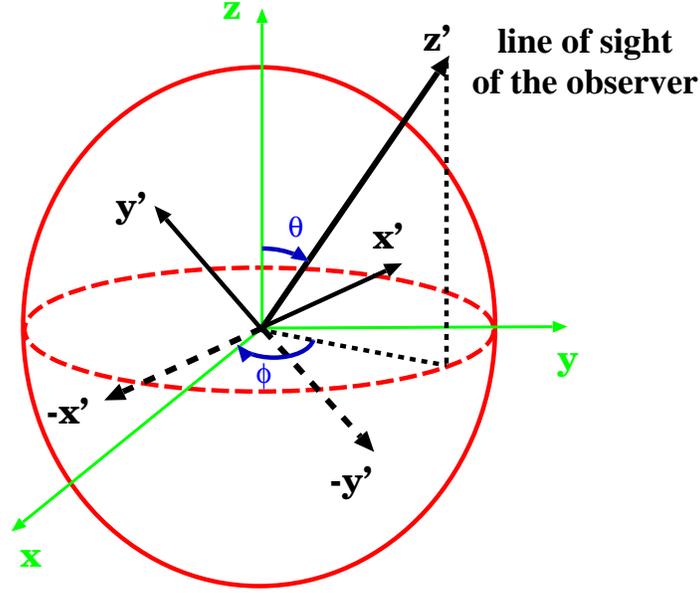}
\caption{ The orientations of the coordinate systems. The Cartesian axes
($x,y,z$) represent the halo principal coordinate system while the axes 
($x',y',z'$) stand for the observers coordinate system with 
$z'$-axis aligned with the line-of-sight direction. The 
$x'$-axis lies in the $x$-$y$ plane. The angle
 $(\theta,\phi)$ represent the polar angle of the line-of-sight
 direction in the $(x,y,z)$-coordinate system. 
\label{fig:frame}}
\end{figure*}

Let $(\theta,\phi)$ be the polar coordinates of the line-of-sight 
direction in the $\vec{x}$-coordinate system. Then the relation 
between the two coordinate systems can be expressed in terms of the 
rotation matrix $A$ \citep{binney85} as 
\begin{equation}
 \vec{x}=A\vec{x'},
\end{equation}
where 
\begin{equation}
A\equiv\left(
\begin{array}{ccc}
 -\sin\phi & -\cos\phi\cos\theta & \cos\phi\sin\theta \\
 \cos\phi  & -\sin\phi\cos\theta & \sin\phi\sin\theta \\
 0         & \sin\theta          & \cos\theta\\
\end{array}
\right) .
\end{equation}
Figure \ref{fig:frame} represents the relative orientation between the 
observer's coordinate system and the halo principal coordinate system. 

\subsection{Density Profile of Triaxial Dark Matter Halos\label{sec:tnfw}}

We adopt the following density profiles of triaxial dark matter halos
proposed by JS02:
\begin{equation}
\label{gnfw}
 \rho(R)=\frac{\delta_{\rm ce}\rho_{\rm crit}(z)}
{(R/R_0)^\alpha(1+R/R_0)^{3-\alpha}},
\end{equation}
where
\begin{equation}
 R^2\equiv c^2\left(\frac{x^2}{a^2}+\frac{y^2}{b^2}
+\frac{z^2}{c^2}\right)\;\;\;(a\leq b\leq c).
\label{rdef}
\end{equation}
The precise value of the inner slope, $\alpha$, is still controversial,
but almost all the N-body simulations based on the collisionless CDM 
scenario indicate values between $1$ and $1.5$
\citep{navarro96,navarro97,moore99,jing00a,fukushige01,fukushige03,power03}.
Thus we consider both $\alpha=1$ and $\alpha=1.5$ below so as to cover a
possible range of the CDM predictions proposed so far.

JS02 defined the concentration parameter in the triaxial model as
\begin{equation}
 c_e\equiv \frac{R_e}{R_0},
\end{equation}
where $R_e$ is chosen so that the mean density within the ellipsoid of
the major axis radius $R_e$ is $\Delta_e\Omega(z)\rho_{\rm crit}(z)$
with\footnote{Note that our definitions 
of $\Delta_{\rm vir}$ and $\Delta_e$ are slightly different
from those of JS02; $\Delta_{\rm vir}({\rm JS02})= 
\Omega(z)\Delta_{\rm vir}$,
and $\Delta_e({\rm JS02})= \Omega(z)\Delta_e$. 
Of course this does not change the definition of $R_e$.}
\begin{equation}
 \Delta_e=5\Delta_{\rm vir}\left(\frac{c^2}{ab}\right)^{0.75}.
\end{equation}
Here $\Omega(z)$ and $\rho_{\rm crit}(z)$ denote the matter density
parameter and the critical density of universe at redshift $z$,
respectively, and $\Delta_{\rm vir}(z)$ denotes the overdensity of
objects virialized at $z$ whose approximate expression is found, e.g.,
in \citet{oguri01}.

Then the characteristic density $\delta_{\rm ce}$ in equation
(\ref{gnfw}) is written in terms of the concentration parameter $c_e$ as
\begin{equation}
 \delta_{\rm ce}=\frac{\Delta_e \Omega(z)}{3}\frac{c_e^3}{m(c_e)},
\end{equation}
where $m(c_e)$ is
\begin{equation}
\label{eq:mce}
 m(c_e)\equiv\frac{c_e^{3-\alpha}}{3-\alpha}
\, {}_2F_1\left(3-\alpha, 3-\alpha; 4-\alpha; -c_e\right),
\end{equation}
with ${}_2F_1\left(a, b; c; x\right)$ being the hypergeometric function.
For $\alpha=1$ and $1.5$, equation (\ref{eq:mce}) simply reduces to
 \begin{eqnarray}
  m(c_e)= \left\{
      \begin{array}{ll}
        \displaystyle{\ln(1+c_e)-\frac{c_e}{1+c_e}} & 
        \mbox{($\alpha=1$)}, \\ 
        \displaystyle{2\ln(\sqrt{c_e}+\sqrt{1+c_e})-2\sqrt{\frac{c_e}{1+c_e}}} &
        \mbox{($\alpha=1.5$)} .
      \end{array}
   \right. 
\end{eqnarray}
Since $R_e$ is empirically related to the (spherical) virial radius 
$r_{\rm vir}$ as $R_e/r_{\rm vir}\simeq 0.45$ 
(JS02), the scaling radius in the triaxial model, $R_0$, for a halo of a
mass $M_{\rm vir}$ is given as
\begin{equation}
 R_0=0.45 \frac{r_{\rm vir}}{c_e}
=\frac{0.45}{c_e} \left(\frac{3M_{\rm vir}}
{4\pi \Delta_{\rm vir}\Omega(z)\rho_{\rm crit}(z)}\right)^{1/3}.
\end{equation} 

Since we do not know the properties of the density profile of an
individual lensing halo, our prediction for the number of arcs is
necessarily statistical in a sense that it should be made after averaging
over appropriate PDFs of the properties of halos. For this purpose, we
adopt the PDFs that JS02 empirically derived from their simulations.
For the axis ratios, they are given as
\begin{equation}
 p(a/c)=\frac{1}{\sqrt{2\pi}\times 0.113}\exp\left[-\frac{\left\{(a/c)(M_{\rm vir}/M_*)^{0.07[\Omega(z)]^{0.7}}-0.54\right\}^2}{2(0.113)^2}\right]\left(\frac{M_{\rm vir}}{M_*}\right)^{0.07[\Omega(z)]^{0.7}},
\label{p_a}
\end{equation}
and 
\begin{equation}
 p(a/b|a/c)=\frac{3}{2(1-\max(a/c,0.5))}\left[1-\left(\frac{2a/b-1-\max(a/c,0.5)}{1-\max(a/c,0.5)}\right)^2\right],
\label{p_ab}
\end{equation}
for $a/b \ge \max(a/c,0.5)$, and $p(a/b|a/c) = 0$ otherwise (JS02). 
Here $M_*$ is the characteristic nonlinear mass so that the rms top-hat
smoothed overdensity at that mass scale is $1.68$. For the concentration
parameter, we adopt
\begin{equation}
 p(c_e)=\frac{1}{\sqrt{2\pi}\times 0.3}
\exp\left[-\frac{(\ln c_e-\ln \bar{c}_e)^2}{2(0.3)^2}\right]\frac{1}{c_e},
\label{p_ce}
\end{equation}
where the fit to the median concentration parameter $\bar{c}_e$ for
$\alpha=1$ is given as \footnote{This expression looks different from
its counterpart (eq. [21]) of JS02 for two reasons. One is due to a typo
in JS02 who omitted the factor $\sqrt{\Delta_{\rm vir}(z_c;{\rm
JS02})/\Delta_{\rm vir}(z;{\rm JS02})}$. Since $\Delta_{\rm vir}({\rm
JS02})= \Omega(z)\Delta_{\rm vir}$ according to the notation of this
paper, this recovers the difference in the latter part.
The other is the fact that we also incorporate the additional 
axis ratio dependence of $\bar{c}_e$ which is noted in 
equation (23) of JS02. This explains the prefactor before 
$A_e$ in equation (\ref{eq:median-ce}) of this paper.}:
\begin{equation}
\label{eq:median-ce}
 \bar{c}_e=1.35\exp\left[-\left\{\frac{0.3}{(a/c)(M_{\rm vir}/M_*)^{0.07[\Omega(z)]^{0.7}}}\right\}^2\right]A_e\sqrt{\frac{\Delta_{\rm vir}(z_c)}{\Delta_{\rm vir}(z)}}\left(\frac{1+z_c}{1+z}\right)^{3/2},
\end{equation}
with $z_c$ being the collapse redshift of the halo of mass $M_{\rm vir}$
(JS02). In the case of $\alpha=1$, we simply use the above expression,
and for $\alpha=1.5$, we use the relation
$\bar{c}_e(\alpha=1.5)=0.5\bar{c}_e(\alpha=1)$ (Keeton \& Madau 2001;
JS02).  JS02 estimated $A_e=1.1$ in the Lambda-dominated CDM model, but
this value is likely to be dependent on the underlying cosmology to some
extent.  As we stressed in Introduction, however, we do not intend to
survey the cosmological parameters but rather focus on the effects of
the properties of the lensing halos. Therefore while we mostly fix the
value $A_e=1.1$, we also vary the value between 0.8 and 1.6 to see its
systematic effect in \S \ref{sec:discussion}.  Incidentally this is
useful in understanding the difference of the predicted number of arcs
between open and Lambda-dominated CDM models found by
\citet{bartelmann98}.

\section{Gravitational Lensing by Triaxial Dark Matter Halos\label{sec:lens}}

In this section, we present several expressions for triaxial dark
matter halos which are useful in calculating gravitational lensing
properties. Under the thin lens approximation, gravitational lensing
properties are fully characterized by the matter density projected along
the line-of-sight \citep*[e.g.,][]{schneider92}. We have to calculate
the mass density profile projected along the arbitrary line-of-sight
directions, because the line-of-sight, in general, does not coincide
with the principal axis of a triaxial dark matter halo. 
 
For simplicity, in this section we redefine $\vec{x}/R_0$ and
$\vec{x'}/R_0$ as $\vec{x}$ and $\vec{x'}$, respectively. In this case,  
\begin{equation}
 \rho(R)=\frac{\delta_{\rm ce}\rho_{\rm crit}(z)}
{R^\alpha(1+R)^{3-\alpha}} ,
\end{equation}
with $R$ being defined by equation (\ref{rdef}).  In terms of the
observer's coordinates $(x',y',z')$, $R$ is written as
\begin{equation}
\label{Robs}
 R = \sqrt{fz'^{2} + gz' + h},
\end{equation}
where
\begin{eqnarray}
f &=& \sin^{2}\theta\left(\frac{c^2}{a^2}\cos^{2}\phi + 
\frac{c^2}{b^2}\sin^{2}\phi\right)
+ \cos^{2}\theta, \label{f}\\
g &=& \sin\theta\sin2\phi\left(\frac{c^2}{b^2}-\frac{c^2}{a^2}\right)x' + 
\sin 2\theta\left(1-\frac{c^2}{a^2}\cos^{2}\phi-
\frac{c^2}{b^2}\sin^{2}\phi\right)y', \label{g}\\
h &=& \left(\frac{c^2}{a^2}\sin^{2}\phi + \frac{c^2}{b^2}\cos^{2}\phi\right)x'^2 + \sin 2\phi\cos\theta\left(\frac{c^2}{a^2}-\frac{c^2}{b^2}\right)x'y' 
+ \left[\cos^{2}\theta\left(\frac{c^2}{a^2}\cos^{2}\phi + 
\frac{c^2}{b^2}\sin^{2}\phi\right) + 
\sin^{2}\theta\right]y'^{2}\label{h}.
\end{eqnarray}
Defining two new variables $z'_*{}^2$ and $\zeta$  
\begin{eqnarray}
z'_*{} &\equiv& \sqrt{f}\left(z' + \frac{g}{2f}\right), \label{newz}\\
\zeta &\equiv& h - \frac{g^{2}}{4f}\label{zeta}, 
\end{eqnarray}
we rewrite equation (\ref{Robs}) as 
\begin{equation}
R = \sqrt{z'_*{}^2 + \zeta^{2}}. 
\end{equation}

Then the convergence $\kappa$ can be expressed as a function of $\zeta$: 
\begin{equation}
 \kappa=\frac{R_0}{\Sigma_{\rm crit}}
\int_{-\infty}^{\infty}\rho(R)dz'
=\frac{R_0}{\Sigma_{\rm crit}}
\int_{-\infty}^{\infty}\frac{1}{\sqrt{f}}\rho
\left(\sqrt{z'_*{}^2+\zeta^2}\right)dz'_* 
\equiv\frac{b_{\rm TNFW}}{2}f_{\rm GNFW}(\zeta),
\label{convergence}
\end{equation}
where 
\begin{equation}
 b_{\rm TNFW} \equiv \frac{1}{\sqrt{f}}
\frac{4\delta_{\rm ce}\rho_{\rm crit}(z)R_0}{\Sigma_{\rm crit}},
\end{equation}
and
\begin{equation}
 f_{\rm GNFW}(r) \equiv 
\int_{0}^{\infty}\frac{1}{\left(\sqrt{r^2+z^2}\right)^\alpha
\left(1+\sqrt{r^2+z^2}\right)^{3-\alpha}}dz. 
\label{f_nfw} 
\end{equation}
The critical surface mass density $\Sigma_{\rm crit}$ is defined by 
\begin{eqnarray}
\Sigma_{\rm crit} \equiv
\frac{c^2 D_{\rm OS}} {4\pi GD_{\rm OL}D_{\rm LS}} ,
\end{eqnarray}
where $D_{\rm OL}$, $D_{\rm OS}$, and $D_{\rm LS}$ denote the angular
diameter distances from the observer to the lens plane, from the
observer to the source plane, and from the lens plane to the source
plane, respectively.

The meaning of the variable $\zeta$ can be easily understood by 
substituting equations (\ref{f})-(\ref{h}) into equation (\ref{zeta}): 
\begin{equation}
\zeta^{2} = \frac{1}{f}\left(Ax'^{2} + Bx'y' + Cy'^{2}\right),
\label{zeta2}
\end{equation}
where 
\begin{eqnarray}
A &\equiv& \cos^{2}\theta\left(\frac{c^2}{a^2}\sin^{2}\phi + 
\frac{c^2}{b^2}\cos^{2}\phi\right) + \frac{c^2}{a^2}\frac{c^2}{b^2}\sin^{2}\theta, 
\label{A} \\
B &\equiv& \cos\theta\sin 2\phi\left(\frac{c^2}{a^2}-\frac{c^2}{b^2}\right), 
\label{B} \\
C &\equiv& \frac{c^2}{b^2}\sin^{2}\phi+\frac{c^2}{a^2}\cos^{2}\phi.
\end{eqnarray}
The quadratic form of equation (\ref{zeta2}) implies that the
iso-$\zeta$ curves are ellipses, and that the position angle of ellipses 
$\psi$ is
\begin{equation}
\psi = \frac{1}{2}\arctan\frac{B}{A-C}.
\label{eqn:psi}
\end{equation}

\begin{figure*}[t]
\epsscale{0.5}
\plotone{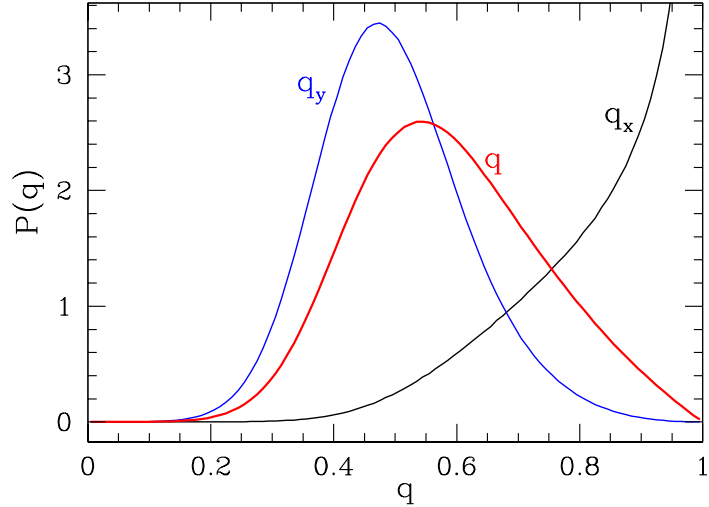}
\caption{PDFs of $q$ (eq. [\ref{eqn:q}]) , $q_x$ (eq.
 [\ref{eqn:qx}]), and $q_y$ (eq. [\ref{eqn:qy}]). Here we consider a halo with
 mass $M_{\rm vir}=10^{15}h^{-1}M_{\odot}$ and redshift $z=0.3$, but the result
 only weakly depends on the halo mass and redshift. These PDFs are calculated
 from PDFs of axis ratios $p(a/c)$ and $p(a/b)$ for which we use
 equations (\ref{p_a}) and (\ref{p_ab}). We assume that the orientations
 of dark halos are random. \label{fig:prob_q}}
\end{figure*}

By rotating the $x'y'$-plane by the angle $\psi$, we diagonalize 
equation (\ref{zeta2}) such that
\begin{equation}
\zeta^{2} = \frac{x'^{2}}{q_{x}^{2}} + \frac{y'^{2}}{q_{y}^{2}}, 
\end{equation}
where
\begin{eqnarray}
q_{x}^{2} &\equiv& \frac{2f}{A + C - \sqrt{(A - C)^{2} + B^{2}}}, 
\label{eqn:qx} \\ 
q_{y}^{2} &\equiv& \frac{2f}{A + C + \sqrt{(A - C)^{2} + B^{2}}}.
\label{eqn:qy} 
\end{eqnarray}
Note that $q_{x} \ge q_{y}$ for the given $\psi$. We further define the 
axis ratio $q$ as
\begin{equation}
q \equiv \frac{q_y}{q_x} = 
\left(\frac{A + C - \sqrt{(A - C)^{2} + B^{2}}}
{A + C + \sqrt{(A - C)^{2} + B^{2}}}\right)^{1/2}, 
\label{eqn:q}
\end{equation}
which represents the ellipticities of the projected
isodensity curves of the triaxial dark halos. 
In this case, the convergence $\kappa$ is expressed as
$\kappa=\kappa(\xi)$, where $\xi^2=x'{}^2+y'{}^2/q^2$. 
The advantage of this diagonalization is that we can apply the previous
method to calculate lensing properties \citep{schramm90,keeton01a} where
the deflection angle $\vec{\beta}=(\beta_{x'},\beta_{y'})$ is expressed
as a one-dimensional integral of the convergence $\kappa(\xi)$:
\begin{eqnarray}
 \beta_{x'}(x',y')  &=& qx'J_0(x',y'),\\
 \beta_{y'}(x',y')  &=& qy'J_1(x',y'),
\end{eqnarray}
where the integral $J_n(x,y)$ is
\begin{equation}
J_n(x,y) = \int_0^1 \frac{\kappa\left(\xi(v)\right)}{\left[1-(1-q^2)v\right]^{n+1/2}}dv, 
\end{equation}
and $\xi(v)$ is
\begin{equation}
  \xi^2(v)=v\left(x^2+\frac{y^2}{1-(1-q^2) v}\right).
\end{equation}
Figure \ref{fig:prob_q} plots PDFs of $q$, $q_{x}$, and
$q_{y}$. They were computed numerically using equations
(\ref{p_a})-(\ref{p_ab}) and (\ref{eqn:qx})-(\ref{eqn:q}) under the
assumption that the triaxial halo orientations (i.e., the angles
$\theta$ and $\phi$) are randomly distributed. In this plot we set
$M_{\rm vir}=10^{15}h^{-1}M_{\odot}$ and $z=0.3$, which are s typical mass
scale and a redshift of lensing clusters. It is clear from 
Figure \ref{fig:prob_q} that the axis ratio of projected isodensity 
contours strongly deviates from unity, having maximum around $q \sim
0.6$. This large degree of ellipticity suggests that the triaxial 
dark halos in realistic cosmological models significantly enhances the 
number of arcs compared with the conventional spherical model
predictions. 

For $\alpha=1$, $f_{\rm GNFW}(r)$ defined in equation (\ref{f_nfw}) 
is analytically expressed as \citep{bartelmann96}:
\begin{eqnarray}
 f_{\rm GNFW}(r)
=\left\{\begin{array}{ll}
{\displaystyle \frac{1}{1-r^2}
\left[-1+\frac{2}{\sqrt{1-r^2}}{\rm arctanh}\sqrt{\frac{1-r}{1+r}}\right]}
&\mbox{($r<1$),}\\
{\displaystyle \frac{1}{r^2-1}
\left[1-\frac{2}{\sqrt{r^2-1}}\arctan\sqrt{\frac{r-1}{r+1}}\right]}
&\mbox{($r>1$),}
\end{array}\right. 
\end{eqnarray}
but it does not has a simple analytical expression for
$\alpha=1.5$. Thus we use the following fitting formula in this case:
\begin{equation}
 f_{\rm GNFW}(r)
=\frac{2.614}{r^{0.5}\left(1+2.378r^{0.5833}+2.617r^{1.5}\right)}.
\label{fit_a15}
\end{equation}
The error of the above fit is $\lesssim 0.6$\%.

\section{Arc Statistics in the Triaxial Dark Matter Halo\label{sec:arcstat}}

\subsection{Cross Sections for Arcs from the 
Monte Carlo Simulation\label{sec:mc}}

First we compute the cross section for arcs without distinguishing
tangential and radial arcs mainly because of the computational cost.  
In fact, the previous analyses indicate that while the number ratio 
of radial to tangential arcs offers another information on the density
profile, the ratio is rather insensitive to the non-sphericity
\citep{molikawa01,oguri01,oguri02a}.

Since the analytical computation of the cross sections is not
practically feasible except for spherical models, we resort to the
direct Monte Carlo method
\citep{bartelmann94,miralda93b,molikawa01,oguri02a}.  We showed that the
convergence of triaxial dark matter halos is expressed by equation
(\ref{convergence}). Thus the corresponding lensing deflection angle
$\vec{\alpha}$, and therefore the cross section $\tilde{\sigma}$, are
fully characterized by the two parameters, $b_{\rm TNFW}$ and $q$, as
long as the finite size of source galaxies is safely neglected. Thus we
perform the Monte Carlo simulations on the dimensionless $X$-$Y$ plane,
where $X$ and $Y$ are $X\equiv x'/(R_0q_x)$ and $Y\equiv y'/(R_0q_x)$,
and tabulate the deflection angle and the dimensionless cross section
\begin{eqnarray} 
 \vec{\alpha}&=&\vec{\alpha}(b_{\rm TNFW},q),\\
\tilde{\sigma}&=&\tilde{\sigma}(b_{\rm TNFW},q) ,
\end{eqnarray}
in $50\times19$ bins ($\alpha=1$) or $70\times19$ bins ($\alpha=1.5$) 
for $b_{\rm TNFW}$ and $q$, respectively. The dimensionless cross section is
translated to the dimensional one in the source plane as
\begin{equation}
 \sigma=\tilde{\sigma}(b_{\rm TNFW},q)
\times \left(R_0q_x\frac{D_{\rm OS}}{D_{\rm OL}}\right)^2 .
\end{equation}

\begin{figure*}[t]
\epsscale{0.6}
\plotone{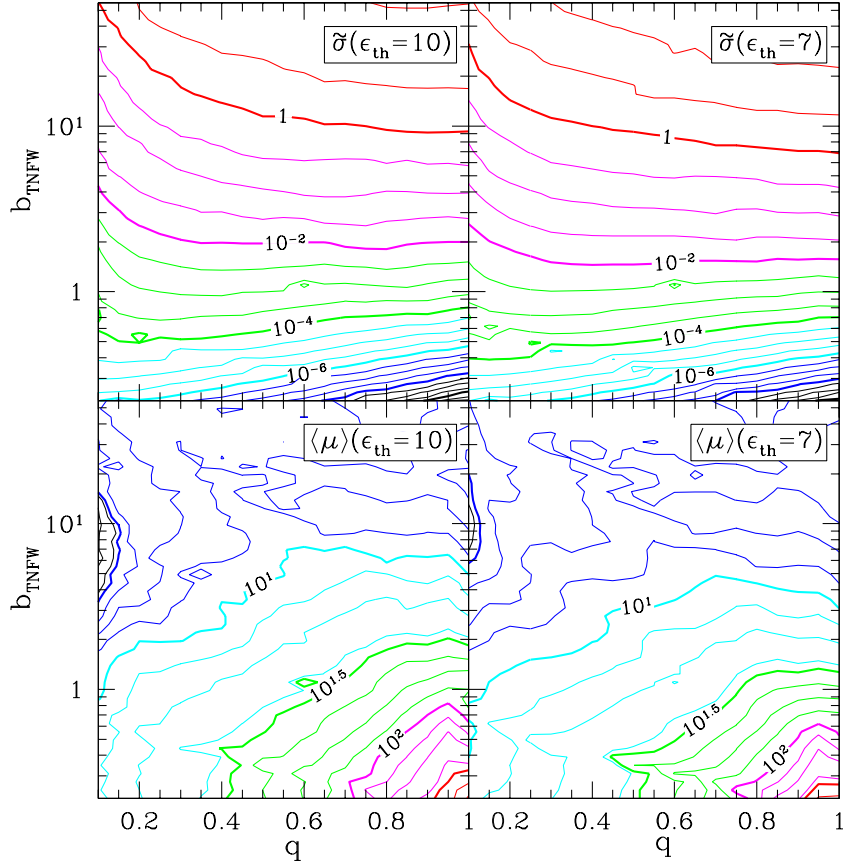}
\caption{ Contours of dimensionless cross sections $\tilde{\sigma}$ and
 average magnification factors $\langle \mu\rangle$ in the 
 $q$-$b_{\rm TNFW}$ plane for $\alpha=1$. The threshold axis ratios for
 arcs are set to $\epsilon_{\rm th}=10$ ({\it upper}) and $7$ ({\it
 lower}), respectively. Contours are drawn at $10^{0.5n}$ for
 $\tilde{\sigma}$ and at $10^{0.125n}$ for  $\langle \mu\rangle$, where
 $n$ is integer. When $n$ is in multiples of $4$, contours are drawn by
 thick lines. These cross sections and magnification factors are derived
 from Monte Carlo simulations described in \S \ref{sec:mc}. 
\label{fig:cross_a10}}
\end{figure*}
\begin{figure*}[t]
\epsscale{0.6}
\plotone{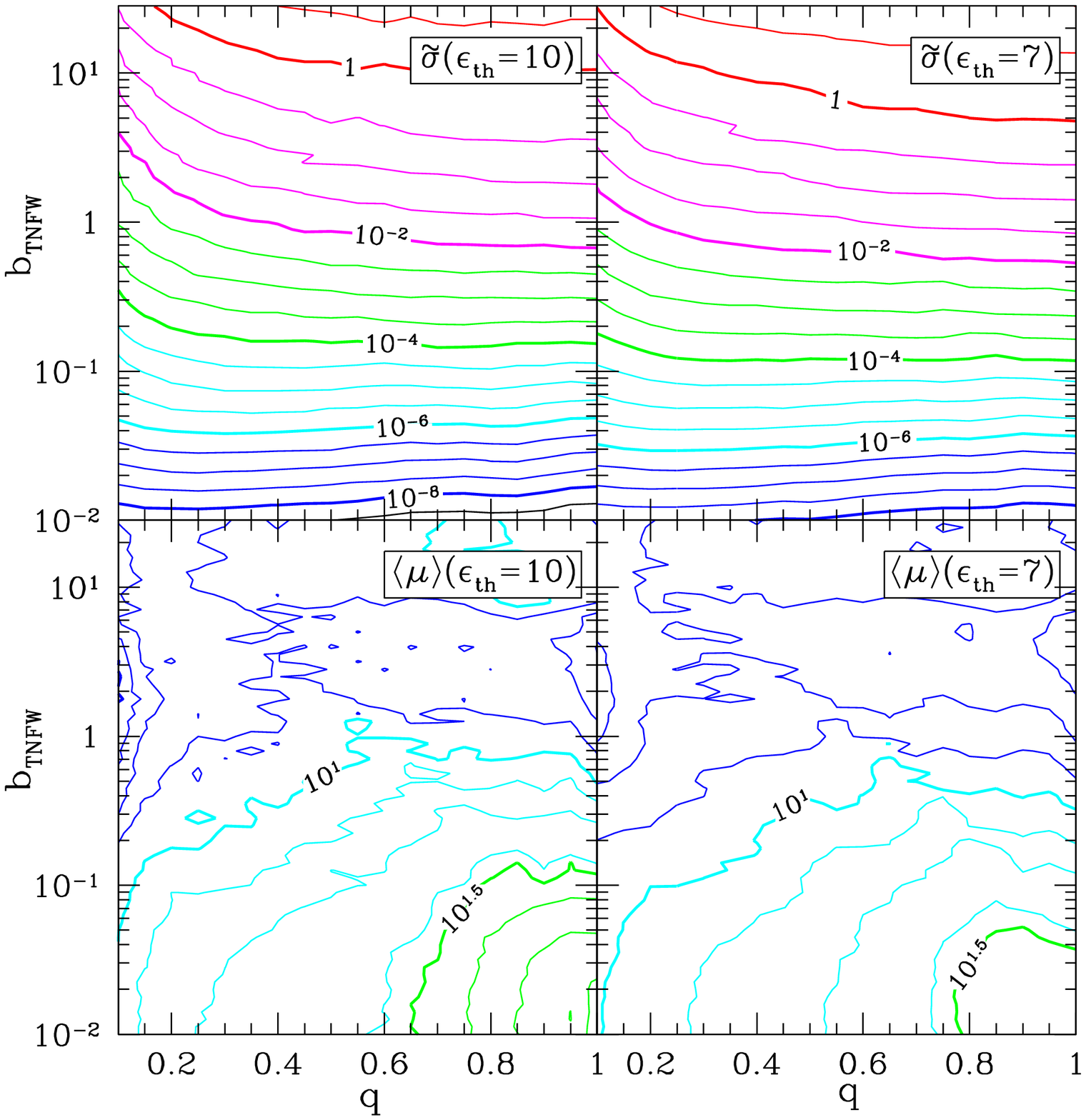}
\caption{Same as Figure \ref{fig:cross_a10}, except for $\alpha=1.5$.
 \label{fig:cross_a15}} 
\end{figure*}

We follow the simulation method by \citet{oguri02a} which is briefly
summarized below. We use a $2048 \times 2048$ regular grid on the
$X$-$Y$ plane and calculate the deflection angle at each grid point.
The box size is adjusted so as to include all arcs in the box for
each $(b_{\rm TNFW}, q)$. Therefore, the box size almost scales as the
tangential critical line for each $(b_{\rm TNFW}, q)$. After those
deflection angles are obtained at each grid point, we trace 
back the corresponding position in the source plane, and see whether or
not it constitutes a part of lensed images. In order to take account of
the source ellipticity which is also important in arc statistics
\citep{keeton01b}, we assume that it distributes randomly in the range
of [0,0.5], where source ellipticity is defined by $1-b_{\rm s}/a_{\rm
s}$ with $a_{\rm s}$ and $b_{\rm s}$ being semi-major and semi-minor
axes, respectively. We adopt this distribution of intrinsic ellipticities
in order to compare our results with the previous works
\citep[e.g.,][]{bartelmann98} in which the same distribution was
assumed. Moreover, the distribution is roughly consistent with the
observed distribution \citep*[e.g.,][]{lambas92}. Once we identify a
lensed image, we compute its length $l$ and width $w$ as described in
\citet{oguri02a}. Finally we define a lensed arc if the ratio of $l$ and
$w$ exceeds the threshold value $\epsilon_{\rm th}$ that we set: 
\begin{equation}
 \frac{l}{w}\geq \epsilon_{\rm th} .
\end{equation}
In practice, we consider $\epsilon_{\rm th}=7$ and $10$ to check the
robustness of the conclusion. We also compute the average magnification
of the arcs $\langle\mu\rangle$ for each set of $(b_{\rm TNFW}, q)$
which is required in estimating the magnification bias
\citep*{turner80,turner84}. The contours of the lensing cross sections
and the average magnification are plotted in Figures
\ref{fig:cross_a10} and \ref{fig:cross_a15} for $\alpha=1$ and
$\alpha=1.5$, respectively.  We confirmed that the cross sections for
$q=1$ cases reproduce the analytic result of spherical lens models for
point source.
 
We should note that our current method does not take account of the
finite size effect of source galaxies, and thus our results are,
strictly speaking, applicable only to a sufficiently small source. Since
the number of tangential arcs, which dominates the total number of arcs,  
is known to be insensitive to the source size
\citep{hattori97b,bartelmann98,molikawa01,oguri01,oguri02a}, this should
not change our conclusion. 

\subsection{Predicting Numbers of Arcs}

The next step is to average the cross section for arcs corresponding to 
a halo of $M_{\rm vir}$ at $z_{\rm L}$ and a galaxy at $z_{\rm S}$ over
the halo properties (its orientations and axis ratios):
\begin{equation}
 \overline{\sigma}(M_{\rm vir}, z_{\rm L}, z_{\rm S})
=\int d(a/c)\int dc_e \int d(a/b) \int d\theta \int d\phi 
~p(a/c)p(c_e)p(a/b|a/c)p(\theta)p(\phi)\sigma.
\label{cs_av}
\end{equation}
In what follows, we assume the orientations of triaxial dark matter halos
are completely random: 
\begin{eqnarray}
 p(\theta)&=&\frac{\sin\theta}{2},\label{p_theta}\\
p(\phi)&=&\frac{1}{2\pi}.
\end{eqnarray}
The realistic prediction for the number of arcs also requires to
properly take account of the magnification bias. Thus we use the
average of the cross section times number density of galaxies above the
magnitude limit:
\begin{equation}
 \overline{\sigma n_{\rm g}}(M_{\rm vir}, z_{\rm L}, z_{\rm S})
=\int d(a/c)\int dc_e \int d(a/b) \int d\theta \int d\phi 
p(a/c) \, p(c_e) \, p(a/b|a/c) \, p(\theta) \, 
p(\phi) \, \sigma \int_{L_{\rm min}}^\infty dL n_g(L,z_{\rm S}),
\label{csng}
\end{equation}
where $n_g(L,z)$ is the luminosity function of source galaxies for which
we adopt the Schechter form:
\begin{equation}
 n_g(L,z)dL =  \phi^*\left(\frac{L}{L^*}\right)^{\alpha_{\rm s}}
\exp\left(-\frac{L}{L^*}\right)\frac{dL}{L^*}.
\label{schechter}
\end{equation} 
Its integral over $L$ simply reduces to
\begin{equation}
 \int_{L_{\rm min}}^\infty dL \, n_g(L,z_{\rm S})
=\phi^*\Gamma(\alpha_{\rm s}+1,L_{\rm min}/L^*),
\end{equation}
with $\Gamma(a,x)$ being the incomplete gamma function of the second
kind.  The lower limit of the integral, $L_{\rm lim}$, may be computed
from limiting magnitude of observation, $m^*$, and the lensing
magnification factor $\langle\mu\rangle$ (see Figs.  \ref{fig:cross_a10}
and \ref{fig:cross_a15}):
\begin{eqnarray}
 \frac{L_{\rm min}}{L^*}
&=& \frac{10^{-0.4(m_{\rm lim}-m^*)}}{\langle\mu\rangle}, \\
 m^* &=& M^*+5\log
\left[\frac{D_{\rm OS}(1+z_{\rm S})^2}{10{\rm pc}}\right]+K(z_{\rm S}) .
\end{eqnarray}
We adopt the K-correction in B-band for spiral galaxies \citep{king85}:
\begin{equation}
K(z)=-0.05+2.35z+2.55z^2-4.89z^3+1.85z^4 .
\label{kcor}
\end{equation}

Finally the number distribution of lensed arcs for a halo of mass
$M_{\rm vir}$ at $z_{\rm L}$ is given by 
\begin{equation}
 \frac{dN_{\rm arc}}{dz_{\rm S}}(z_{\rm S};M_{\rm vir},z_{\rm L})=\overline{\sigma n_{\rm g}}(M_{\rm vir}, z_{\rm L}, z_{\rm S})\frac{cdt}{dz_{\rm S}}(1+z_{\rm S})^3 ,
\label{dnds}
\end{equation}
and the total number of lensed arcs for the halo is
\begin{equation}
 N_{\rm arc}(M_{\rm vir},z_{\rm L})
=\int_{z_{\rm L}}^{z_{\rm S,max}} dz_{\rm S}\,
\frac{dN_{\rm arc}}{dz_{\rm S}}(z_{\rm S};M_{\rm vir},z_{\rm L}).
\label{n_arc}
\end{equation}
While the upper limit of redshifts of source galaxies, $z_{\rm S,max}$,
is in principle arbitrary, it is practically limited by the validity of
the input luminosity function of source galaxies and the applied
K-correction at high redshifts. In the present analysis, we
conservatively set $z_{\rm S,max}=1.25$ because of the K-correction (eq.
[\ref{kcor}]) and the luminosity function (\S \ref{sec:lf}).
Nevertheless we stress here that our methodology can be applied to at
higher redshifts if they are replaced by any reliable models valid
there.
 
\subsection{Luminosity Function of Source Galaxies\label{sec:lf}}

While the predicted number of arcs sensitively depends on the luminosity
function of source galaxies \citep[e.g.,][]{hamana97}, $n_{\rm g}(L.z)$
is still fairly uncertain especially at high $z$. Thus we consider the
following four luminosity functions measured up to $z=1.25$: HDF1 from
the Hubble Deep Field and the New Technology Telescope Deep Field
\citep{poli01}, HDF2 from the Hubble Deep Field \citep*{sawicki97}, SDF
from the Subaru Deep Field \citep{kashikawa03}, and CFRS from the
Canada-France Redshift Survey \citep{lilly95}. They are summarized
in Table \ref{table:lf}. Although the Schechter fits to those luminosity
functions are valid only at $z> (0.2 \sim 0.6)$, we simply extrapolate
the values even down to $z=0$ if necessary. This does not affect our
result in \S \ref{sec:obs} at all since galaxies at $z\sim 1$ are the
main sources of lensed arcs for our sample of clusters at $z>0.2$ (\S
\ref{sec:clusteremss}). 

Except for HDF1, the Schechter parameters were derived assuming the
Einstein-de Sitter (EdS) model ($\Omega_0=1$, $\lambda_0=0$) in the
original references. We convert them into the counterparts in the
Lambda-dominated universe ($\Omega_0=0.3$, $\lambda_0=0.7$) as follows.

Since the number of galaxies in the redshift interval $[z_{\rm S},z_{\rm
S}+dz_{\rm S}]$,
\begin{equation}
 dN_{\rm g}(z_{\rm S})\propto D^2_{\rm OS}\frac{c\,dt}{dz_{\rm S}}dz_{\rm S}n_{\rm g}(L,z_{\rm S})dL, 
\end{equation}
is observable, it should be invariant. Thus the luminosity function in
the Lambda-dominated universe is related to that in the EdS as:
\begin{equation}
 \left[n_{\rm g}(L',z_{\rm S})dL'\right]_{\rm Lambda}
=\frac{\left[D^2_{\rm OS}(c\,dt/dz_{\rm S})\right]_{\rm EdS}}
{\left[D^2_{\rm OS}(c\,dt/dz_{\rm S})\right]_{\rm Lambda}}
\left[n_{\rm g}(L,z_{\rm S})dL\right]_{\rm EdS}, 
\end{equation}
where 
\begin{equation}
 L' \equiv \frac{\left[D^2_{\rm OS}\right]_{\rm Lambda}}
{\left[D^2_{\rm OS}\right]_{\rm EdS}}L.
\end{equation}

\begin{figure*}[t]
\epsscale{0.6}
\plotone{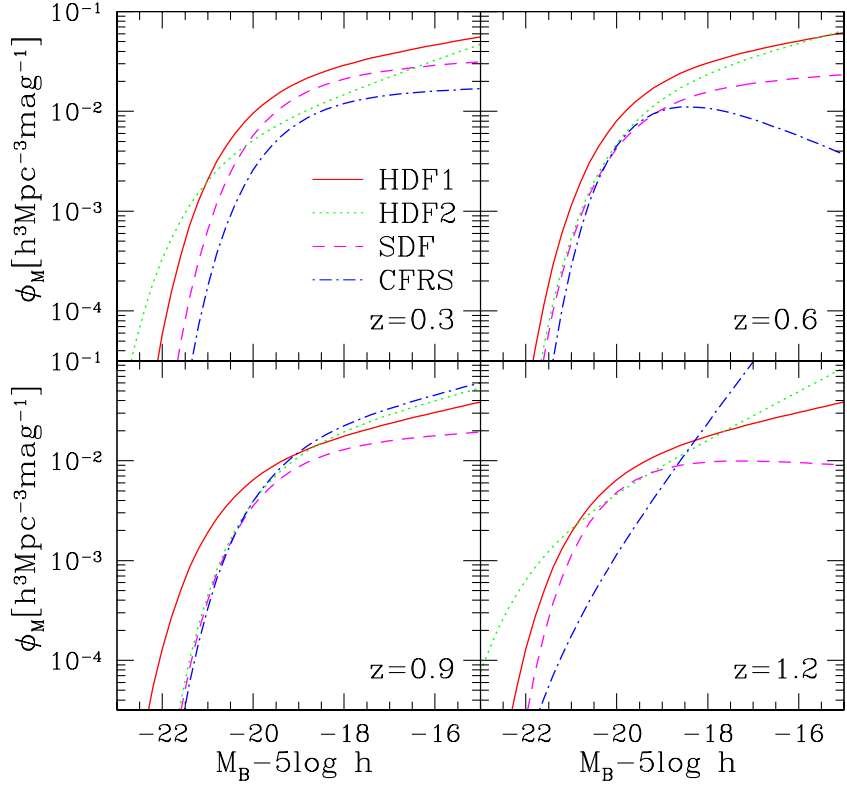}
\caption{Luminosity functions of source galaxies (eq.
 [\ref{schechter_m}]) for $z=0.3$, $0.6$,  $0.9$, and $1.2$. Parameters
 of these luminosity functions are  summarized in Table \ref{table:lf}.
 \label{fig:lf}} 
\end{figure*}

The resulting luminosity functions in terms of the absolute
magnitude $M$:
\begin{equation}
 \phi_M(M,z)dM=0.921\phi^*10^{-0.4(\alpha_{\rm s}+1)(M-M^*)}
\exp\left(-10^{-0.4(M-M^*)}\right)dM,
\label{schechter_m}
\end{equation}
at $z=0.3$, $0.6$, $0.9$, and $1.2$ are plotted in Figure \ref{fig:lf}.
Clearly the uncertainty increases at fainter luminosities at $z>1$,
which may significantly change the predicted number of arcs.  Therefore,
while we adopt HDF1 as our fiducial model, we also attempt to evaluate
the uncertainty due to the different choice of luminosity functions using
the other three. 

\subsection{Predicted Cross Sections and Numbers of Arcs}

\begin{figure*}[t]
\epsscale{0.6}
\plotone{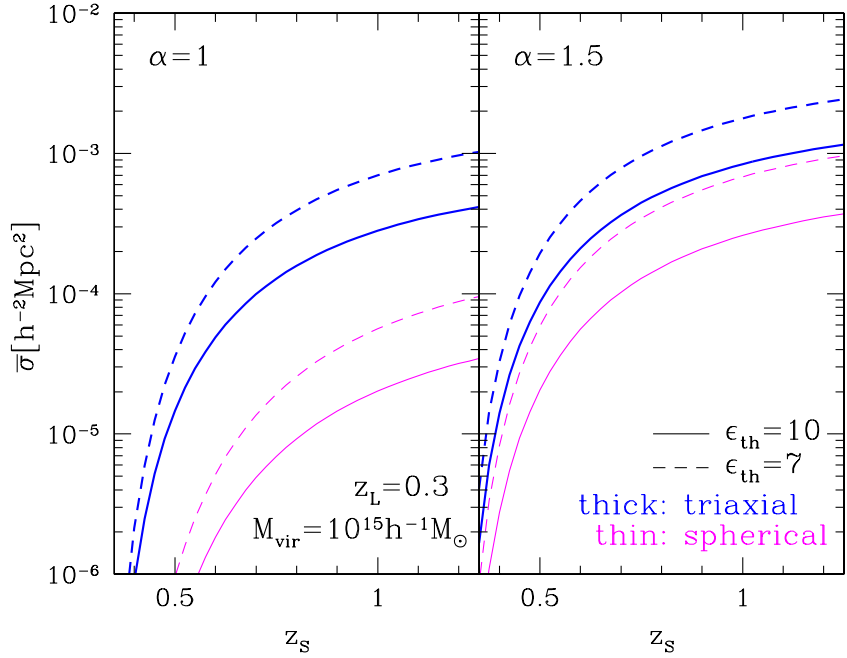}
\caption{ Average cross sections (eq. [\ref{cs_av}]) for triaxial and
 spherical dark matter halo models as a function of source redshift
 $z_{\rm S}$ for both $\alpha=1$ ({\it left}) and $1.5$ ({\it right}),
 where $\alpha$ is the inner slope of dark matter halo density profile. 
The lens cluster has a mass 
$M_{\rm vir}=10^{15}h^{-1}M_\odot$ and is placed at $z_{\rm L}=0.3$.
For the threshold axis ratio of arcs, we adopt both
 $\epsilon_{\rm th}=10$ ({\it solid}) and $7$ ({\it dashed}). 
\label{fig:cross_zs}} 
\end{figure*}
\begin{figure*}[t]
\epsscale{0.6}
\plotone{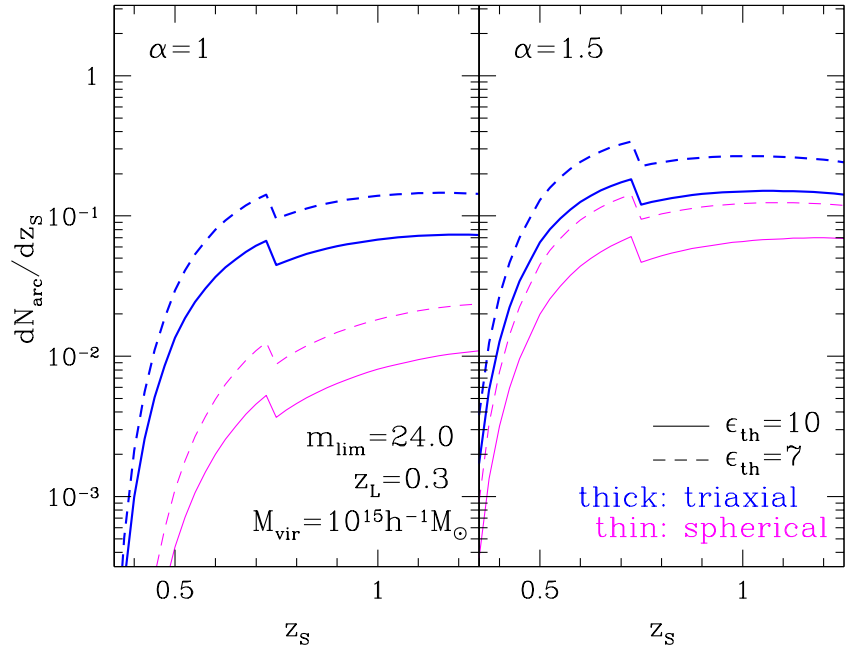}
\caption{Number distributions of arcs (eq. [\ref{dnds}]) for triaxial and
 spherical dark matter halo models. The B-band magnitude limit for
 arcs is set to $m_{\rm lim}=24$. The distributions are discontinuous
 at $z_{\rm S}=0.75$ because we adopt binned luminosity function (see
 Table \ref{table:lf}).
\label{fig:dndzs}} 
\end{figure*}

\begin{figure*}[t]
\epsscale{0.6}
\plotone{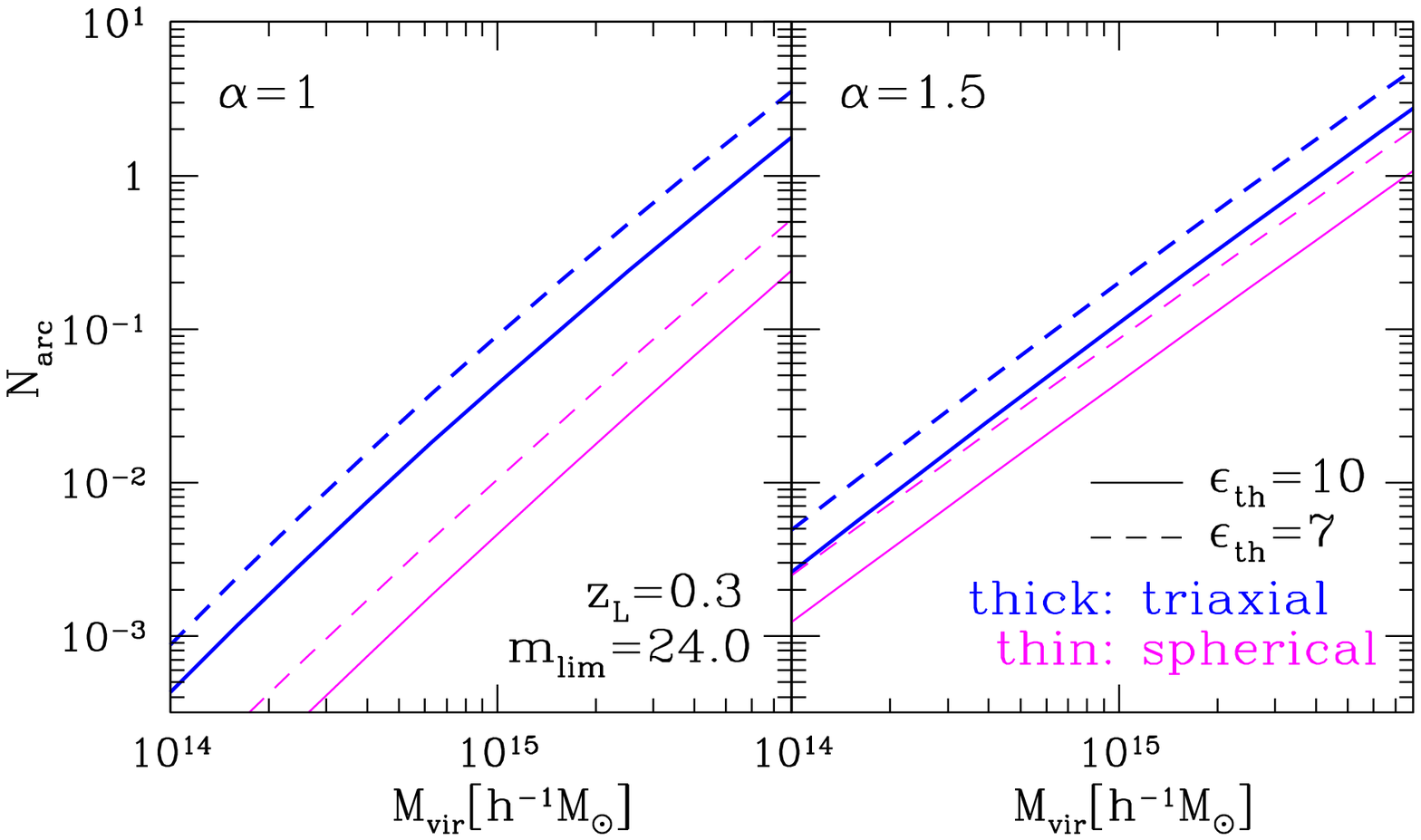}
\caption{Predicted numbers of arcs (eq. [\ref{n_arc}]) as a function of
 halo mass $M_{\rm vir}$. The redshift of the dark halo is still fixed
 to $z_{\rm L}=0.3$. The B-band magnitude limit for
 arcs is set to $m_{\rm lim}=24$. 
\label{fig:num_mvir}} 
\end{figure*}

Figure \ref{fig:cross_zs} shows the average cross sections (eq.
[\ref{cs_av}]) of a dark matter halo of $M_{\rm
vir}=10^{15}h^{-1}M_\odot$ at $z_{\rm L}=0.3$.  The cross section for
the triaxial model is larger by a factor of 10 ($\alpha=1$) and of 4
($\alpha=1.5$) than that for the spherical counterpart.  Since the
magnification factor is always larger for smaller cross sections (see
Figures \ref{fig:cross_a10} and \ref{fig:cross_a15}), the magnification
bias further reduces the difference between $\alpha=1$ and 1.5 for the
triaxial model. This explains the behavior of Figure \ref{fig:dndzs}
where the source redshift distribution of arcs (eq. [\ref{dnds}]) is
plotted. Actually the figure indicates that the non-spherical effect
even exceeds that of the difference due to the inner slope.

\begin{figure*}[t]
\epsscale{0.6}
\plotone{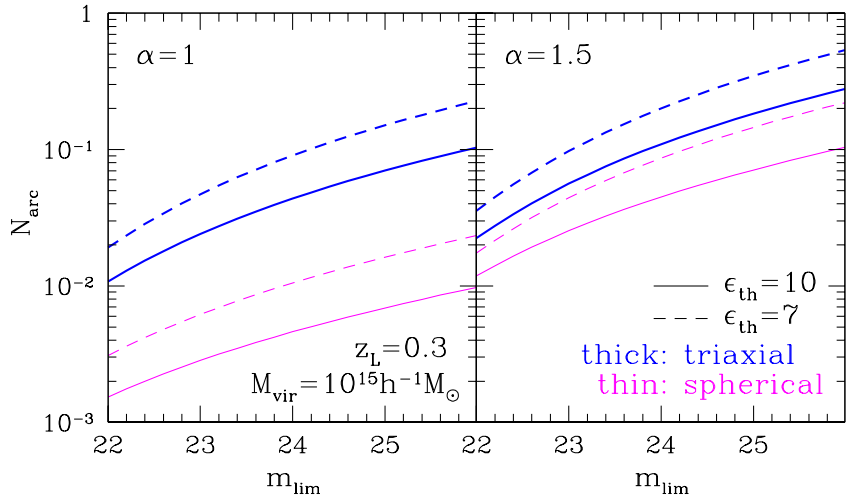}
\caption{Predicted numbers of arcs as a function of B-band magnitude limit
 $m_{\rm lim}$. The mass of lens cluster is $M_{\rm vir}=10^{15}h^{-1}M_\odot$. 
\label{fig:num_mlim}} 
\end{figure*}

\begin{figure*}[t]
\epsscale{0.6}
\plotone{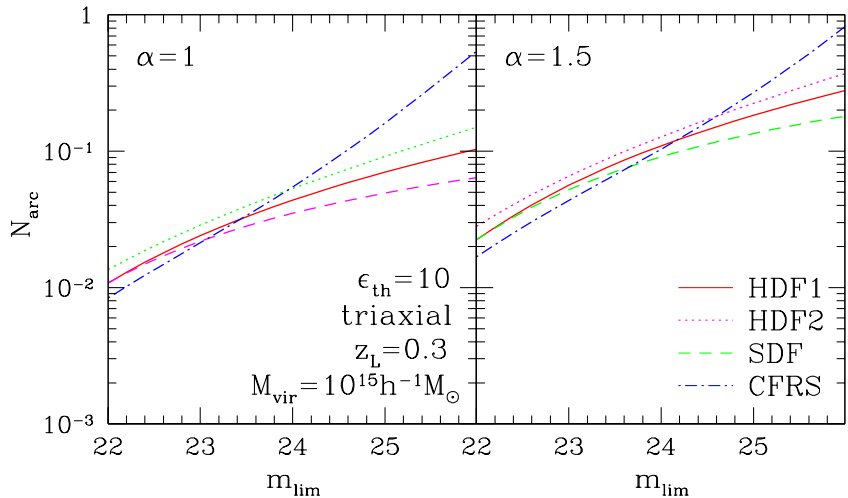}
\caption{Predicted numbers of arcs for different luminosity functions of
 source galaxies as a function of B-band magnitude limit
 $m_{\rm lim}$. Parameters of luminosity functions are given in Table
 \ref{table:lf}. Only triaxial dark matter halo model and threshold axis
 ratio $\epsilon_{\rm th}=10$ are considered.
\label{fig:num_mlim_lf}} 
\end{figure*}

Figures \ref{fig:num_mvir}, \ref{fig:num_mlim} and 
\ref{fig:num_mlim_lf} show how the
predicted number of arcs depends on the mass of a lensing halo, the
limiting magnitude of the survey, and the adopted luminosity function of
source galaxies.  Figure \ref{fig:num_mvir} shows that the number of
arcs is sensitive to the mass of halo, implying the estimate of the mass
of the target cluster is essential in interpreting the data. In
addition, the difference between $\alpha=1$ and 1.5 becomes smaller for
the triaxial model of $M_{\rm vir}>10^{15}M_\odot$. Thus in order to
distinguish the inner slope clearly as well, one needs a sample of less
massive clusters that have lensed arcs.

Figure \ref{fig:num_mlim} indicates that the number of arcs is also
sensitive to the magnitude limit, suggesting that the well-controlled
selection function for the arc survey is quite important.  On the other
hand, the uncertainty of the luminosity function of source galaxies
seems to be less critical, at least for arcs of galaxies at $z_{\rm
s}<1.25$ that we consider in this paper (Fig. \ref{fig:num_mlim_lf}).
The difference among the four luminosity functions (see Table
\ref{table:lf}) is merely up to 50 \% for $m_{\rm lim}< 24$, and is
within a factor of 2 even at $m_{\rm lim}< 26$ except CFRS.  The
predictions based on HDF1 approximately correspond to the median among
the four and this is why we choose this as our fiducial model in what
follows.

\section{Comparison with the Observed Number of Arcs\label{sec:obs}}

\subsection{Cluster Data\label{sec:clusteremss}}

We use a sample of $38$ X-ray selected clusters compiled by
\citet{luppino99}. The clusters are selected from the {\it Einstein
Observatory} Extended Medium Sensitivity Survey (EMSS). For all the
clusters, deep imaging observations with B-band limiting magnitude
$m_{\rm lim}\sim 26.0$ were carried out to search for arcs.

As we remarked in the previous section, the mass estimate of those
clusters is important in understanding the implications from the
observed arcs statistics. For this purpose, we first construct a gas
temperature -- X-ray luminosity (in the Einstein band) relation from a
subset of the above clusters whose temperature is determined.  Then we
estimate the temperature of the remaining clusters using the temperature
-- luminosity relation. Finally we estimate the mass of each cluster
employing the virial mass -- gas temperature relation of
\citet*{finoguenov01}. 

\begin{figure*}[t]
\epsscale{0.5}
\plotone{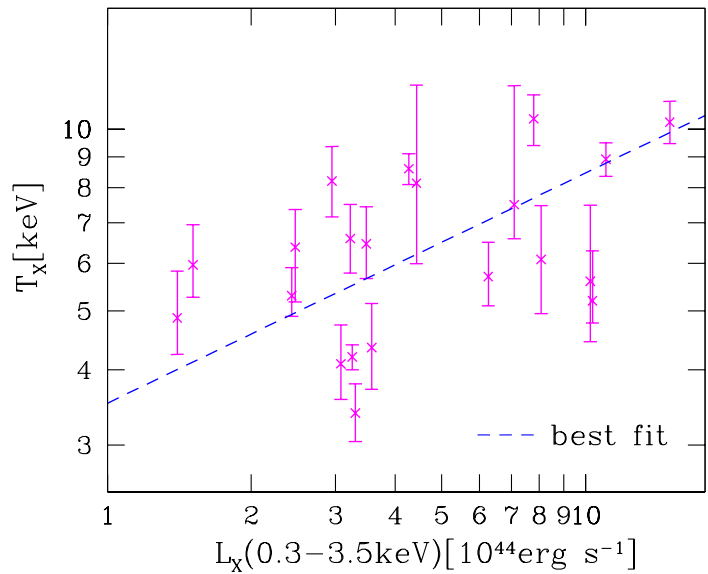}
\caption{The luminosity -- temperature relation for the EMSS cluster
 sample. Among 38 clusters, we use 21 clusters with measured temperature
 to derive luminosity -- temperature relation. The best-fit
 luminosity -- temperature relation is shown in equation (\ref{ltrelation}).
\label{fig:lt}} 
\end{figure*}

More specifically, our best-fit luminosity -- temperature relation from
Figure \ref{fig:lt} is
\begin{equation}
 T_X=T_{X,0}\left(\frac{L_X(0.3-3.5{\rm keV})}
{10^{44}{\rm erg\, s^{-1}}}\right)^\gamma,
\label{ltrelation}
\end{equation}
where $\gamma=0.381\pm 0.052$ and $T_{X,0}=3.52^{+0.32}_{-0.29}{\rm
keV}$. The derived luminosity -- temperature relation is consistent with
recent other estimations \citep[e.g.,][]{ikebe02}.
Neglecting the possible redshift evolution for the
luminosity -- temperature relation \citep[e.g.,][]{mushotzky97}, we
estimate the temperature of those clusters without spectroscopic data as
shown in Table \ref{table:emss}. The mass -- temperature relation that we
adopt is
\begin{equation}
 T_X=2.3{\rm keV}\left(\frac{M_{\rm vir}}{10^{14}h^{-1}M_\odot}\right)^{0.54}.
\end{equation}
This relation is derived by \citet{shimizu03} who converted the result
of \citet{finoguenov01} in terms of $M_{\rm vir}$ assuming the density
profile (eq. [\ref{gnfw}]; the difference between $\alpha=1$ and 1.5
turned out to be negligible).

\subsection{Observed Number of Arcs\label{sec:obs_num}}

The observed giant arcs ($\epsilon_{\rm th}=10$) in the $38$ EMSS
cluster sample are listed in Table \ref{table:arc}. The number of arcs
in this sample is roughly consistent with more recent data from
different cluster samples \citep{zaritsky03,gladders03}.  In order to be
consistent with our adopted luminosity functions and K-correction of
source galaxies, we need to select the arcs with $z<1.25$.  In
reality, this is quite difficult; most of the observed arcs do not have
a measured redshift, while uncertainties of source redshifts may
systematically change lensing probabilities. For instance,
\citet{wambsganss03} explicitly showed that it is important to take
correctly account of the source redshift which can change cross sections
by an order of magnitude.   
Moreover four in the list labeled ``Candidate'' in Table \ref{table:arc}
are even controversial and may not be real lensed arcs. Thus we consider
the two extreme cases; one is to select only the two arcs with measured
redshifts less than 1.25, and the other is to assume that all the arcs
without measured redshifts in the list (including the candidates) are
located at $z<1.25$. Of course the reality should be somewhere in
between, and thus we assume that the range between the two cases well
represents the current observational error. This means that the
observational error can be greatly reduced if redshifts of all arcs are
measured in the future observations. 

\subsection{Comparison of Theoretical Predictions with Observations}

\begin{figure*}[t]
\epsscale{0.6}
\plotone{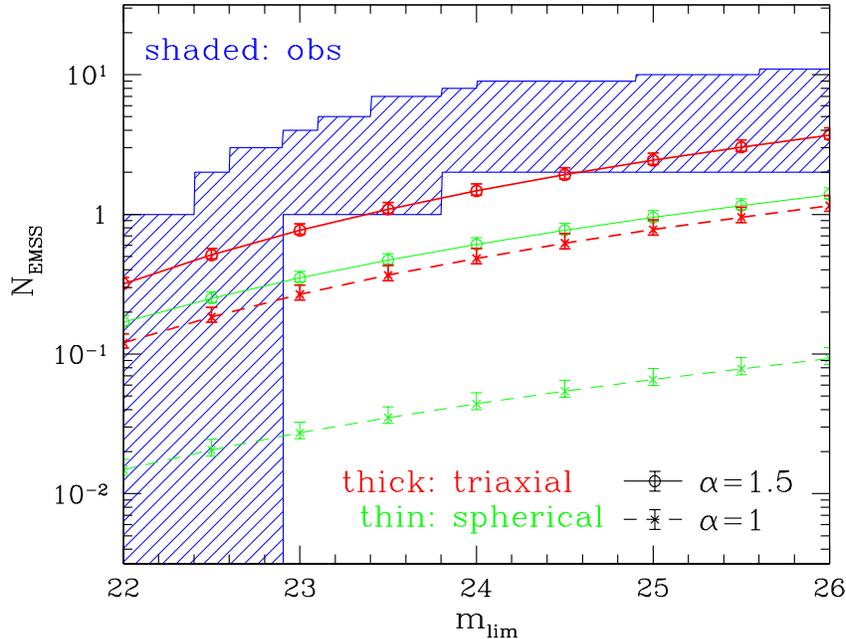}
\caption{The number of arcs in the 38 EMSS cluster sample (eq.
 [\ref{n_emss}]) as a function of B-band limiting magnitude $m_{\rm lim}$. 
 The threshold axis ratio is $\epsilon_{\rm th}=10$. The observed number 
 of arcs taking account of several uncertainties, which is shown by the
 shaded region, is discussed in \S \ref{sec:obs_num}. 
\label{fig:num_obs_mlim}} 
\end{figure*}

Finally let us compare our theoretical predictions with the data in
detail. Our prediction of the number of arcs is the sum of equation
(\ref{n_arc}) over all the 38 EMSS clusters:
\begin{equation}
 N_{\rm EMSS}\equiv\sum^{38}_{i=1}N_{\rm arc}(M_{{\rm vir},i},z_{{\rm L},i}).
\label{n_emss}
\end{equation}
We also compute the error of the predicted number of arcs by propagating
the mass uncertainty for each cluster (Table \ref{table:emss}).  Figure
\ref{fig:num_obs_mlim} shows the number of arcs in the 38 EMSS cluster
sample as a function of the B-band limiting magnitude $m_{\rm lim}$.
When the B-band magnitude of an arc is not available, we convert its
corresponding V- or R-band magnitude into the B-band assuming typical
colors of spiral galaxies at $z\sim 1$, $B-V=V-R=1$ \citep{fukugita95}.

The important conclusion that we draw from Figure \ref{fig:num_obs_mlim}
is that the triaxial model in the Lambda-dominated CDM universe with the
inner slope of $\alpha=1.5$ successfully reproduces the observed number of
arcs, and that the spherical model prediction with $\alpha=1$ fails by a
wide margin. Both the triaxial model with $\alpha=1$ and the spherical
model with $\alpha=1.5$ are marginal in a sense that the presence of
substructure in the dark halo which we ignore in the current method
should systematically increase our predicted number of arcs. Indeed
\citet*{meneghetti03a} reported that the substructure enhances the number
of arcs with $\epsilon_{\rm th}=10$ typically by a factor 2 or 3. This
is exactly the amount of enhancement that is required to reconcile those
two models with the observation.

We note here that the additional contribution due to galaxies inside a
cluster is generally small; \citet*{flores00} and \citet{meneghetti00}
found that galaxies increase the number of arcs merely by $\sim$10\%.
Even a central cD galaxy produces the number of arcs by not more than
$\sim$50\% \citep*{meneghetti03b}.

\section{Discussion\label{sec:discussion}}

\subsection{Comparison with the previous result}

Our result that the halos in a Lambda-dominated CDM universe reproduces
the observed number of arcs seems inconsistent with the previous result of
\citet{bartelmann98} who claimed that only open CDM models can reproduce
the observation. One possibility to explain the apparent discrepancy is
the difference of the inner profile of halos; we showed that the slope
of $\alpha=1.5$ is required to reproduce the observation.  This implies
that N-body simulations may underestimate the real number of arcs unless
they have sufficient spatial resolution. On the other hand,
cluster-scale halos may indeed have a shallower inner profile
\citep{jing00a}. Therefore this is closely related to the well known
problem of the inner slope of CDM dark matter halos
\citep{navarro96,navarro97,moore99,jing00a,fukushige01,fukushige03,power03},
and would need further investigation.  Moreover, in reality, the mass
estimate for each cluster, the limiting magnitude of source galaxies,
and the adopted luminosity function would also affect the prediction in
a more complicated fashion, and the further quantitative comparison is
not easy at this point.

\begin{figure*}[t]
\epsscale{0.6}
\plotone{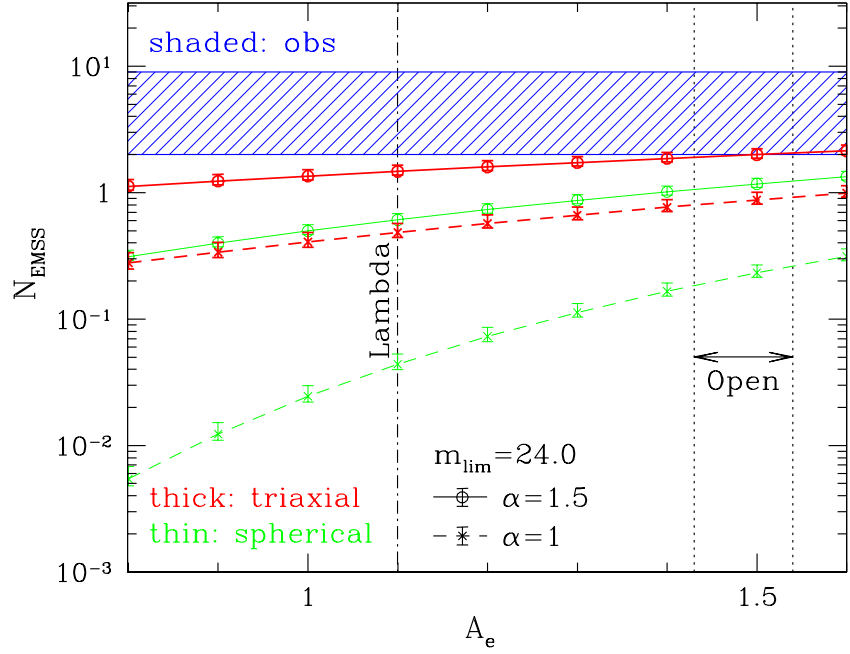}
\caption{The number of arcs in the 38 EMSS cluster sample as a function
 of $A_e$ for $m_{\rm lim}=24$. Dash-dotted line indicates the fiducial value
 for $A_e$, $A_e=1.1$, in a Lambda-dominated CDM model. Dotted lines suggest
 possible range of $A_e$ with taking account of the enhancement of the
 concentration parameter in an open CDM model (see text for details).
\label{fig:num_obs_ce}} 
\end{figure*}

Nevertheless we can point out the general tendency that open CDM models
produce more arcs than Lambda-dominated CDM models because of the larger
value of the concentration parameter in the former. Thus it is unlikely that
difference between open and Lambda-dominated CDM models results from the
``global'' effect of the cosmological parameters.  In order to show this, we
compute the number of arcs as a function of  $A_e$ still assuming the
Lambda-dominated CDM model. Figure \ref{fig:num_obs_ce} plots $N_{\rm
EMSS}$ for $m_{\rm lim}=24$ as a function of $A_e$. While JS02 found
$A_e=1.1$ in a Lambda-dominated CDM models, their fitting formula \citep[see
also][]{bartelmann98} tend to predict $\sim 30-40\%$ larger
concentration parameter in open CDM models. This enhancement of the
concentration parameter corresponds to $A_e=1.43\sim1.53$ if we still
assume Lambda-dominated CDM models as a background cosmology. Thus the
effect of $A_e$ alone increases the number of arc by $\sim 50\%-100\%$
even for triaxial cases, which is qualitatively consistent with the
result of \citet{bartelmann98}.

\subsection{Required Non-sphericity of Lensing Halos}

\begin{figure*}[t]
\epsscale{0.6}
\plotone{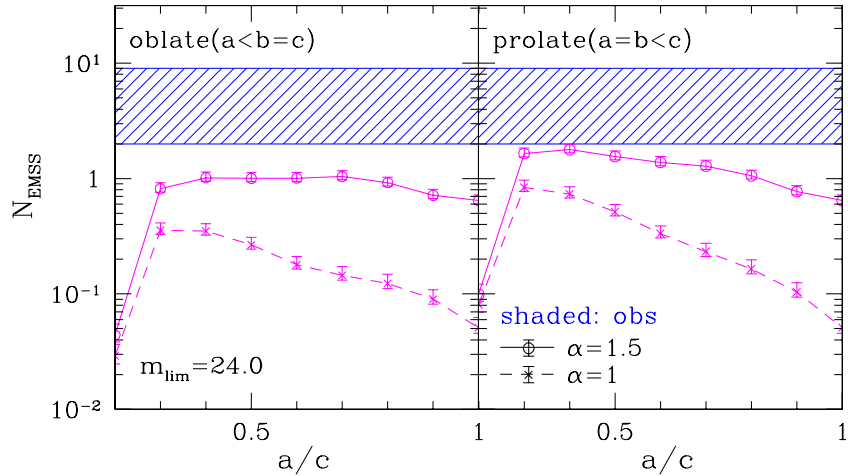}
\caption{The number of arcs in the 38 EMSS cluster sample for fixed axis
 ratios of dark matter halos. The B-band limiting magnitude is set to
 $m_{\rm lim}=24$. Left panel plots the oblate case ($a<b=c$) while
 right panel is the prolate case ($a=b<c$).
\label{fig:num_obs_def}} 
\end{figure*}

Although we showed that the triaxial halos predicted in the
Lambda-dominated CDM model
reproduce the observed number of arcs, the analysis employed a series of 
fairly complicated PDFs for the axial ratios of JS02, and it
is not so clear what degree of non-sphericity for lensing halos is
required to account for the observation.  Thus we rather simplify the
situation and consider that all halos consist of oblate ($a<b=c$) or
prolate ($a=b<c$) halos with a fixed axial ratio. This is equivalent to
replacing $p(a/c)$ (eq. [\ref{p_a}]) or $p(a/b)$ (eq. [\ref{p_ab}]) by
the corresponding $\delta$-functions.  Figure \ref{fig:num_obs_def}
plots the result of this exercise.

The predicted number of arcs is indeed sensitive to the axis ratios of
dark matter halos, and prolate halos of $a/c\lesssim0.5$ in the
$\alpha=1.5$ case reproduce the observation. This is basically
consistent with the finding of JS02 for halo properties.

The reason why prolate halos tend to produce the larger number of arcs
than oblate halos is explained as follows.  Notice first that to keep
the mass of dark matter halo invariant with the change of the axial
ratio, $b_{\rm TNFW}$ should be approximately proportional to
$(ab/c^2)^{-1}$. Suppose that oblate and prolate halos are projected
onto their axisymmetric direction ($x$ for oblate and $z$ for
prolate). Then their lensing cross sections should scale as
\begin{eqnarray}
 \sigma({\rm oblate})
&\propto & \tilde{\sigma}((a/c)(a/c)^{-1}b_{\rm TNFW},1)
=\tilde{\sigma}(b_{\rm TNFW},1),\\
 \sigma({\rm prolate})
&\propto & \left(\frac{a}{c}\right)^2\tilde{\sigma}((a/c)^{-2}b_{\rm TNFW},1) 
=\left(\frac{a}{c}\right)^{2-2\delta}\tilde{\sigma}(b_{\rm TNFW},1),
\end{eqnarray}
where we assume $\tilde{\sigma}(b_{\rm TNFW},q)\propto b_{\rm
TNFW}^\delta$.  Since Figures \ref{fig:cross_a10} and
\ref{fig:cross_a15} suggest $\delta\gtrsim 2$, we find that $\sigma({\rm
prolate})\gg\sigma({\rm oblate})$ for $a/c <1$.  If those halos are
projected along the $y$-direction, on the other hand, their cross
sections are almost the same:
\begin{equation}
 \sigma({\rm oblate})\sim\sigma({\rm prolate})
\propto \left(\frac{a}{c}\right)^{-\delta}\tilde{\sigma}(b_{\rm TNFW},a/c).
\end{equation}
The above consideration explains the qualitative difference between
oblate and prolate halos, and points out that the elongation along the
line-of-sight is also important in the arc statistics as well as the
asymmetry of the projected mass density.

\subsection{Are Clusters Equilibrium Dark Matter Halos?}

So far we have assumed the one-to-one correspondence between dark
matter halos and X-ray clusters.  This assumption, however, is
definitely over-simplified \citep{suto01,suto03}.  If ``dark clusters''
which are often reported from recent weak lensing analyses \citep{hattori97a,wittman01,miyazaki02} are real, the one-to-one correspondence
approximation may be unexpectedly inaccurate.  As an extreme
possibility, let us suppose that observed X-ray clusters preferentially
correspond to halos in equilibrium.  According to \citet{jing00b}, such
halos have generally larger concentration parameters and their scatter
is small.  In order to imitate this situation, we repeat the computation
using $A_e=1.3$ and the scatter of $0.18$ (Jing 2000; JS02).  We
find that this modified model increases the number of arcs merely
by 10\%$-$20\%. Thus our conclusion remains the same.

\subsection{Sample Variance}

\begin{figure*}[t]
\epsscale{0.6}
\plotone{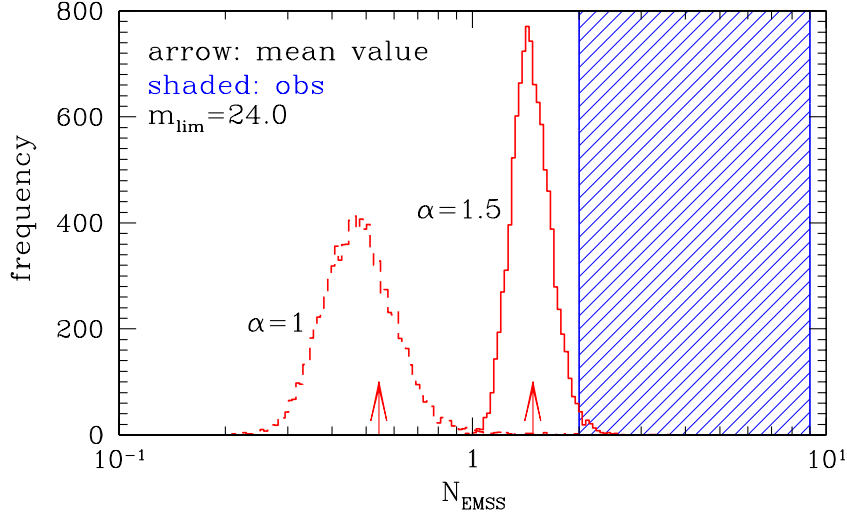}
\caption{The effect of the sample variance. The number of arcs in the 38
 EMSS cluster sample is calculated, not being averaged over axis ratios,
 orientation, and concentration parameters, but using fixed values for
 each clusters.  Axis ratios, orientation, and concentration parameters
 for each cluster are randomly chosen according to their corresponding 
 PDFs (\S \ref{sec:tnfw}).  We calculate 10000 realizations and plot the
 histogram of frequency. Averaged values are shown by arrows. 
\label{fig:samvar}} 
\end{figure*}

The predicted number of arcs for the EMSS cluster that we have presented
so far is based on the {\it averaged} cross section. This is a
reasonably good approximation in the situation that the number of sample
clusters is large enough, but in the current sample, its validity is not
clear.  To examine the sample variance, we re-compute the number of arcs
in the 38 EMSS cluster sample without using the average statistics.
Instead, we first randomly choose values of the axis ratios, the
orientation angles, and the concentration parameters for each cluster
according to their corresponding PDFs (\S
\ref{sec:tnfw}). Then we sum up the number of arcs for the entire
cluster sample.  We repeat the procedure 10000 times each for $\alpha=1$
and 1.5, and construct a distribution function of $N_{\rm EMSS}$ as
plotted in Figure \ref{fig:samvar}. The resulting 1$\sigma$ sample
variance is $\sim$30\% for $\alpha=1$ and $\sim$15\% for $\alpha=1.5$.  
Therefore we confirm that the effect of the sample variance does not 
change our overall conclusion.

\section{Conclusions \label{sec:conclusion}}

We have presented a semi-analytic method to predict the number of lensed
arcs, for the first time taking proper account of the triaxiality of
lensing halos. We found that Lambda-dominated CDM models successfully
reproduce the observed number of arcs of X-ray-selected clusters
\citep{luppino99} if the inner slope of the density profile is close to
$\alpha=1.5$.  Since the spherical models significantly underestimate
the expected number of arcs, we conclude that the observed number of
arcs indeed requires the non-sphericity of the lensing halos.  In fact,
the number of arcs is sensitive to the axis ratios of those halos, and
the non-sphericity that reproduces the observed number corresponds to the
minor to major axis ratios of $\sim 0.5$. This value is perfectly
consistent with the findings of JS02 in the Lambda-dominated CDM models.
In this sense, we may even argue that the arc statistics lend strong
support for the collisionless CDM paradigm at the mass scale of
clusters.  As discussed in \citet{meneghetti01}, self-interacting dark
matter models \citep{spergel00} for instance, are inconsistent with the
observed number of arcs not only because they erase the central cusp but
because they produce much rounder dark matter halos
\citep{yoshida00a,yoshida00b}. Since we have exhibited that even the
current arc surveys have a great impact in testing the collisionless CDM
paradigm, larger surveys with well-controlled systematics in near future
will unveil the nature of dark matter more precisely. 

{\bf Note added} -- In the published version, we used an incorrect PDF
of the angle $\theta$ (eq. [47]). In this version of the preprint, we
used the correct PDF and replaced all related figures to the correct
ones. In these new plots, however, the number of arcs in the triaxial
dark matter halo model becomes smaller only by $\sim 30$\%, so this does
not affect the conclusion of the paper.

\acknowledgments

We thank Y.~P. Jing for useful correspondences concerning many aspects
of the triaxial dark matter halo model from N-body simulations, and
Masahiro Takada for discussions.  We also thank an anonymous referee for
many useful comments. J. L. acknowledges gratefully the research grant
of the JSPS (Japan Society of Promotion of Science) fellowship.  This
research was supported in part by the Grant-in-Aid for Scientific
Research of JSPS (12640231). 


\begin{deluxetable}{lccccccl}
\tablewidth{0pt}
 \tablecaption{B-band luminosity functions of source galaxies used in
 this paper.\label{table:lf}}
\tablehead{\colhead{Name} & \colhead{Model} & \colhead{$z$ Range} &
 \colhead{$\alpha_{\rm s}$} &
 \colhead{$M^*_{AB}-5\log h$\tablenotemark{a}} & \colhead{$\phi^*[h^3{\rm Mpc^{-3}]}$}& \colhead{Ref.} }
\startdata
 HDF1 & Lambda\tablenotemark{b} & $0.00$ -
 $0.50$\tablenotemark{c} & $-1.19$ & $-20.26$ & $2.5 \times 10^{-2}$ & 1\\
 & & $0.50$ - $0.75$ & $-1.19$ & $-19.97$ & $2.9 \times 10^{-2}$ & \\
 & & $0.75$ - $1.25$ & $-1.25$ & $-20.61$ & $1.2 \times 10^{-2}$ & \\
 HDF2 & EdS\tablenotemark{d} & $0.00$ -
 $0.50$\tablenotemark{c} & $-1.40$ & $-21.20$ & $9.0 \times 10^{-3}$ & 2\\
 & & $0.50$ - $1.00$ & $-1.30$ & $-19.90$ & $4.2 \times 10^{-2}$ & \\
 & & $1.00$ - $1.25$ & $-1.60$ & $-22.10$ & $6.0 \times 10^{-3}$ & \\
 SDF & EdS\tablenotemark{d} & $0.00$ -
 $1.00$\tablenotemark{c} & $-1.07$ & $-19.78$ & $4.2 \times 10^{-2}$ & 3\\
 & & $1.00$ - $1.25$ & $-0.92$ & $-20.13$ & $4.3 \times 10^{-2}$ & \\
 CFRS & EdS\tablenotemark{d} & $0.00$ -
 $0.50$\tablenotemark{c} & $-1.03$ & $-19.53$ & $2.7 \times 10^{-2}$ & 4\\
 & & $0.50$ - $0.75$ & $-0.50$ & $-19.32$ & $6.2 \times 10^{-2}$ & \\
 & & $0.75$ - $1.00$ & $-1.28$ & $-19.73$ & $5.4 \times 10^{-2}$ & \\
 & & $1.00$ - $1.25$\tablenotemark{f} & $-2.50$ & $-21.36$ & $9.6 \times 10^{-4}$ & \\
\enddata
\tablenotetext{a}{B-band AB magnitude can be converted to conventional
 Johnson-Morgan magnitude via $B_{AB}=B-0.14$ \citep*{fukugita95}.} 
\tablenotetext{b}{$\Omega_0=0.3$, $\lambda_0=0.7$.} 
\tablenotetext{c}{Extrapolated to $z=0$.}
\tablenotetext{d}{$\Omega_0=1$, $\lambda_0=0$.} 
\tablenotetext{e}{The luminosity function for blue galaxies only.}
\tablerefs{(1) \citealt{poli01}; (3) \citealt{sawicki97}; (3)
 \citealt{kashikawa03}; (4) \citealt{lilly95}}
\end{deluxetable}
\begin{deluxetable}{lcccccl}
\tablewidth{0pt}
 \tablecaption{Properties of clusters in the $38$ EMSS distant cluster
 sample.\label{table:emss}}
\tablecolumns{2}
\tablehead{& & 
 \colhead{$L_X$(EdS)\tablenotemark{a}} &
 \colhead{$L_X$(Lambda)\tablenotemark{b}} & 
 \colhead{$T_X$} & 
 \colhead{$M_{\rm vir}$} & \\ 
 \colhead{Name} & \colhead{$z_{\rm L}$} &
 \colhead{[$10^{44}{\rm erg\, s^{-1}}$]} &
 \colhead{[$10^{44}{\rm erg\, s^{-1}}$]} &
 \colhead{[${\rm keV}$]} & \colhead{[$10^{14}h^{-1}M_\odot$]} & 
 \colhead{Ref.} } 
\startdata
MS 0011.7$+$0837 & $0.163$ &  $3.77$ & $2.24$ & $4.79^{+0.48}_{-0.44}$\tablenotemark{c} & $3.89^{+0.72}_{-0.66}$ & 1 \\
MS 0015.9$+$1609 & $0.546$ & $14.64$ &$11.03$ & $8.92^{+0.57}_{-0.56}$ & $12.30^{+1.46}_{-1.43}$ & 1, 2\\
MS 0302.5$+$1717 & $0.425$ &  $2.88$ & $2.04$ & $4.62^{+0.45}_{-0.41}$\tablenotemark{c} & $3.64^{+0.66}_{-0.60}$ & 1 \\
MS 0302.7$+$1658 & $0.426$ &  $5.04$ & $3.57$ & $4.35^{+0.80}_{-0.64}$ & $3.25^{+1.11}_{-0.89}$ & 1, 2\\
MS 0353.6$-$3642 & $0.320$ &  $5.24$ & $3.48$ & $6.46^{+0.98}_{-0.80}$ & $6.77^{+1.90}_{-1.55}$ & 1, 2\\
MS 0433.9$+$0957 & $0.159$ &  $4.34$ & $2.57$ & $5.04^{+0.53}_{-0.48}$\tablenotemark{c} & $4.27^{+0.83}_{-0.75}$ & 1 \\
MS 0440.5$+$0204 & $0.190$ &  $4.01$ & $2.43$ & $5.30^{+0.60}_{-0.40}$ & $4.69^{+0.98}_{-0.66}$ & 1, 3\\
MS 0451.5$+$0250 & $0.202$ &  $6.98$ & $4.27$ & $8.60^{+0.50}_{-0.50}$ & $11.50^{+1.24}_{-1.24}$ & 1, 3\\
MS 0451.6$-$0305 & $0.539$ & $19.98$ &$15.00$ & $10.27^{+0.85}_{-0.80}$ & $15.97^{+2.45}_{-2.30}$ & 1, 2\\
MS 0735.6$+$7421 & $0.216$ &  $6.12$ & $3.79$ & $5.85^{+0.68}_{-0.61}$\tablenotemark{c} & $5.63^{+1.21}_{-1.09}$ & 1 \\
MS 0811.6$+$6301 & $0.312$ &  $2.10$ & $1.40$ & $4.87^{+0.95}_{-0.63}$ & $4.01^{+1.45}_{-0.96}$ & 1, 2\\
MS 0839.8$+$2938 & $0.194$ &  $5.35$ & $3.25$ & $4.20^{+0.20}_{-0.20}$ & $3.05^{+0.27}_{-0.27}$ & 1, 3\\
MS 0906.5$+$1110 & $0.180$ &  $5.77$ & $3.47$ & $5.65^{+0.64}_{-0.58}$\tablenotemark{c} & $5.28^{+1.11}_{-1.00}$ & 1 \\
MS 1006.0$+$1201 & $0.221$ &  $4.82$ & $2.99$ & $5.34^{+0.58}_{-0.52}$\tablenotemark{c} & $4.76^{+0.96}_{-0.86}$ & 1 \\
MS 1008.1$-$1224 & $0.301$ &  $4.49$ & $2.95$ & $8.21^{+1.15}_{-1.05}$ & $10.55^{+2.74}_{-2.50}$ & 1, 2\\
MS 1054.5$-$0321 & $0.823$ &  $9.28$ & $7.79$ & $10.4^{+1.00}_{-1.00}$ & $16.35^{+2.91}_{-2.91}$ & 1, 4\\
MS 1137.5$+$6625 & $0.782$ &  $7.56$ & $6.26$ & $5.70^{+0.80}_{-0.60}$ & $5.37^{+1.40}_{-1.05}$ & 1, 5\\
MS 1147.3$+$1103 & $0.303$ &  $2.30$ & $1.51$ & $5.96^{+0.99}_{-0.69}$ & $5.83^{+1.79}_{-1.25}$ & 1, 2\\
MS 1201.5$+$2824 & $0.167$ &  $2.03$ & $1.21$ & $3.78^{+0.34}_{-0.32}$\tablenotemark{c} & $2.51^{+0.42}_{-0.39}$ & 1 \\
MS 1208.7$+$3928 & $0.340$ &  $2.03$ & $1.37$ & $3.97^{+0.36}_{-0.33}$\tablenotemark{c} & $2.75^{+0.46}_{-0.42}$ & 1 \\
MS 1224.7$+$2007 & $0.327$ &  $4.61$ & $3.08$ & $4.09^{+0.65}_{-0.52}$ & $2.90^{+0.85}_{-0.68}$ & 1, 2\\
MS 1231.3$+$1542 & $0.238$ &  $2.88$ & $1.81$ & $4.41^{+0.42}_{-0.39}$\tablenotemark{c} & $3.34^{+0.59}_{-0.55}$ & 1 \\
MS 1241.5$+$1710 & $0.549$ & $10.70$ & $8.07$ & $6.09^{+1.38}_{-1.14}$ & $6.07^{+2.55}_{-2.10}$ & 1, 2\\
MS 1244.2$+$7114 & $0.225$ &  $3.84$ & $2.39$ & $4.90^{+0.50}_{-0.46}$\tablenotemark{c} & $4.06^{+0.77}_{-0.71}$ & 1 \\
MS 1253.9$+$0456 & $0.230$ &  $3.14$ & $1.96$ & $4.55^{+0.44}_{-0.40}$\tablenotemark{c} & $3.54^{+0.63}_{-0.58}$ & 1 \\
MS 1358.4$+$6245 & $0.327$ & $10.62$ & $7.09$ & $7.50^{+4.30}_{-0.91}$\tablenotemark{d} & $8.93^{+9.48}_{-2.01}$ & 1, 2, 6\\
MS 1426.4$+$0158 & $0.320$ &  $3.71$ & $2.47$ & $6.38^{+0.98}_{-1.20}$ & $6.62^{+1.88}_{-2.30}$ & 1, 2\\
MS 1455.0$+$2232 & $0.259$ & $16.03$ &$10.23$ & $5.60^{+1.88}_{-1.15}$\tablenotemark{d} & $5.20^{+3.23}_{-1.98}$ & 1, 3, 6\\
MS 1512.4$+$3647 & $0.372$ &  $4.81$ & $3.30$ & $3.39^{+0.40}_{-0.35}$ & $2.05^{+0.45}_{-0.39}$ & 1, 2\\
MS 1546.8$+$1132 & $0.226$ &  $2.94$ & $1.83$ & $4.43^{+0.43}_{-0.39}$\tablenotemark{c} & $3.37^{+0.61}_{-0.55}$ & 1 \\
MS 1618.9$+$2552 & $0.161$ &  $2.24$ & $1.33$ & $3.92^{+0.36}_{-0.33}$\tablenotemark{c} & $2.68^{+0.46}_{-0.42}$ & 1 \\
MS 1621.5$+$2640 & $0.426$ &  $4.55$ & $3.22$ & $6.59^{+0.92}_{-0.81}$ & $7.02^{+1.82}_{-1.60}$ & 1, 2\\
MS 1910.5$+$6736 & $0.246$ &  $4.39$ & $2.78$ & $5.20^{+0.55}_{-0.50}$\tablenotemark{c} & $4.53^{+0.89}_{-0.81}$ & 1 \\
MS 2053.7$-$0449 & $0.583$ &  $5.78$ & $4.43$ & $8.14^{+3.68}_{-2.15}$ & $10.39^{+8.70}_{-5.08}$ & 1, 2\\
MS 2137.3$-$2353 & $0.313$ & $15.62$ &$10.34$ & $5.20^{+1.09}_{-0.42}$\tablenotemark{d}  & $4.53^{+1.76}_{-0.68}$ & 1, 2, 6\\
MS 2255.7$+$2039 & $0.288$ &  $2.04$ & $1.33$ & $3.92^{+0.36}_{-0.33}$\tablenotemark{c} & $2.68^{+0.46}_{-0.42}$ & 1 \\
MS 2301.3$+$1506 & $0.247$ &  $3.29$ & $2.08$ & $4.65^{+0.46}_{-0.42}$\tablenotemark{c} & $3.68^{+0.67}_{-0.62}$ & 1 \\
MS 2318.7$-$2328 & $0.187$ &  $6.84$ & $4.14$ & $6.05^{+0.73}_{-0.65}$\tablenotemark{c} & $6.00^{+1.34}_{-1.19}$ & 1 \\
\enddata
\tablecomments{Errors are at $68$\% confidence limit.}
\tablenotetext{a}{X-ray luminosity in the $0.3-3.5{\rm keV}$ band for
 $\Omega_0=1$, $\lambda_0=0$, and $h=0.5$ universe.} 
\tablenotetext{b}{X-ray luminosity in the $0.3-3.5{\rm keV}$ band for
 $\Omega_0=0.3$, $\lambda_0=0.7$, and $h=0.7$ universe.} 
\tablenotetext{c}{Estimated from $L_X-T_X$ relation (eq.
 [\ref{ltrelation}]).}   
\tablenotetext{d}{The effects of cooling flows are corrected.}  
\tablerefs{(1) \citealt{luppino99}; (2) \citealt*{novicki02}; (3)
 \citealt{mushotzky97}; (4) \citealt{jeltema01}; (5)
 \citealt{borgani01}; (6) \citealt{allen98}}
\end{deluxetable}
\begin{deluxetable}{lccccccl}
\tablewidth{0pt}
 \tablecaption{Giant arcs ($l/w>10$) in the $38$ EMSS distant cluster sample.\label{table:arc}}
\tablehead{\colhead{Cluster} & \colhead{$z_{\rm L}$} & \colhead{Arc} &
 \colhead{$z_{\rm S}$} & \colhead{$l/w$} & \colhead{$m_{\rm arc}$} &
 \colhead{Notes} & \colhead{Ref.} } 
\startdata
MS 0302.7$+$1658 & $0.426$ & A1 & $\sim 0.8$\tablenotemark{a} & $>18$ &
 $B=23.8$ & \nodata & 1, 2\\
               &           & A1W& \nodata & $>12$ & $B=24.9$ & \nodata & \\
MS 0440.5$+$0204 & $0.190$ & A1 & $0.532$ & $>10$ & $B=22.9$ & \nodata &
 1, 3, 4\\
                 &         & A3 & \nodata & $>20$ & $B=24.0$ & \nodata & \\
MS 0451.6$-$0305 & $0.539$ & A1 & \nodata & $10$  & $V=24.6$ & \nodata & 1\\
MS 1006.0$+$1201 & $0.221$ & A2$+$A3  & \nodata & $>20$& $V<22.1$ &
 Candidate & 1, 5\\
                 &         & A4 & \nodata & $12.9$ & $V=21.4$ & Candidate & \\
MS 1008.1$-$1224 & $0.301$ & A2 & \nodata & $10.0$ & $B=23.4$ & Candidate & 1\\
MS 1358.4$+$6245 & $0.328$ & A1 & $4.92$  & $>21$  & \nodata  & \nodata
 & 1, 6\\
MS 1621.5$+$2640 & $0.426$ & A1 & \nodata & $>18$  & $B=23.1$ & \nodata
 & 1, 7\\
MS 1910.5$+$6736 & $0.246$ & A1 & \nodata & $10.5$ & $R=20.6$ &
 Candidate & 1, 5\\
MS 2053.7$-$0449 & $0.583$ & AB & \nodata & $>22$  & $V=22.4$ & \nodata
 & 1, 7\\
MS 2137.3$-$2353 & $0.313$ & A1 & $1.501$ & $18.1$ & $B=22.0$ & \nodata
 & 1, 8, 9, 10\\
\enddata
\tablenotetext{a}{Estimated from color of the arc.} 
\tablerefs{(1) \citealt{luppino99}; (2) \citealt{mathez92}; (3)
 \citealt{luppino93}; (4) \citealt{gioia98} (5); \citealt{lefevre94};
 (6) \citealt{franx97}; (7) \citealt{luppino92}; (8) \citealt{fort92};
 (9) \citealt{hammer97}; (10) \citealt{sand02}}
\end{deluxetable}
\end{document}